\documentclass[a4paper,11pt]{article}
\usepackage{jheppub} 
\usepackage{physics}
\usepackage{slashed}
\usepackage{caption}
\usepackage{xcolor}
\usepackage{comment}
\usepackage{multirow}
\usepackage{graphics}
\usepackage{float}
\usepackage{cancel}
\usepackage{soul}
\usepackage{cases}
\usepackage{array}
\usepackage{mathtools}  
\usepackage{amsfonts}
\usepackage{hyperref}
\usepackage{amsmath}
\usepackage{amssymb}
\usepackage{tcolorbox}
\usepackage{tikz} 
\usepackage{tikz-cd}
\usepackage{tcolorbox}
\usepackage{booktabs}
\usepackage{mdframed}
\usepackage{tikz}
\usepackage{xcolor}
\usepackage{amsmath}
\usepackage{amssymb}
\usepackage{mathrsfs}
\usepackage[utf8]{inputenc}
\usepackage{textgreek}
\usetikzlibrary{arrows.meta, positioning}

\definecolor{TwistorPurple}{RGB}{88,36,130} 
\definecolor{TwistorBlue}{RGB}{36,72,130} 
\definecolor{WittenGreen}{RGB}{0,100,100}
\definecolor{PenroseGray}{RGB}{80,80,80}
\definecolor{penroseblue}{RGB}{30, 60, 90}
\definecolor{penrosegrey}{RGB}{200, 200, 200}

\newenvironment{3dBMSNeq1}
{%
\mdfsetup{%
    skipabove=12pt,
    skipbelow=12pt,
    innertopmargin=1.2\baselineskip,
    innerbottommargin=1\baselineskip,
    innerleftmargin=1em,
    innerrightmargin=1em,
    linecolor=penroseblue,
    backgroundcolor=penroseblue!10,
    linewidth=1pt,
    roundcorner=10pt,
    frametitleaboveskip=0pt,
    frametitlealignment=\raggedright,
    frametitlefont=\bfseries\color{white},
    frametitlebackgroundcolor=penroseblue,
    frametitle={\strut The $\mathcal{N}=1$ supersymmetric $\mathfrak{bms}_3$ algebra}
}
\begin{mdframed}
}
{\end{mdframed}}

\newenvironment{3dBMSNextended}
{%
\mdfsetup{%
    skipabove=12pt,
    skipbelow=12pt,
    innertopmargin=1.2\baselineskip,
    innerbottommargin=1\baselineskip,
    innerleftmargin=1em,
    innerrightmargin=1em,
    linecolor=penroseblue,
    backgroundcolor=WittenGreen!10,
    linewidth=1pt,
    roundcorner=10pt,
    frametitleaboveskip=0pt,
    frametitlealignment=\raggedright,
    frametitlefont=\bfseries\color{white},
    frametitlebackgroundcolor=WittenGreen,
    frametitle={\strut The  $\mathcal{N}-$extended supersymmetric $\mathfrak{bms}_3$ algebra}
}
\begin{mdframed}
}
{\end{mdframed}}

\newenvironment{4dBMSNsusy}
{%
\mdfsetup{%
    skipabove=12pt,
    skipbelow=12pt,
    innertopmargin=1.2\baselineskip,
    innerbottommargin=1\baselineskip,
    innerleftmargin=1em,
    innerrightmargin=1em,
    linecolor=penroseblue,
    backgroundcolor=penroseblue!15,
    linewidth=1pt,
    roundcorner=10pt,
    frametitleaboveskip=0pt,
    frametitlealignment=\raggedright,
    frametitlefont=\bfseries\color{white},
    frametitlebackgroundcolor=penroseblue,
    frametitle={\strut The $\mathcal{N}-$extended supersymmetric $\mathfrak{bms}_4$ algebra}
}
\begin{mdframed}
}
{\end{mdframed}}

\title{\boldmath\boldmath Light-Ray Supersymmetry, BMS Algebras and the Averaged Null Energy Condition}

\author{Dhruva K.S.}

\affiliation{Department of Theoretical Physics, Tata Institute of Fundamental Research,\\
Homi Bhabha Rd, Mumbai 400005, India}

\emailAdd{dhruvacaesar@gmail.com}

\abstract{We show that the light-ray algebra of unitary supersymmetric Lorentzian conformal field theories  contains a universal supersymmetric generalized $\mathfrak{bms}$ algebra. We construct parallel light-ray operators using the supersymmetry current, $R-$symmetry current and the stress tensor supported on a co-dimension one null hypersurface and determine their graded commutators in $d=3,4$ to verify this claim. Our construction only relies on global supersymmetry and fundamental physical principles such as microcausality, unitarity and Lorentz invariance and thus applies to strongly coupled theories. As a corollary, we show that the 
averaged null energy operator is a positive operator using the local supersymmetry algebra on the null sheet, providing a supersymmetric derivation of the averaged null energy condition (ANEC). Finally, we verify our results in free massless supersymmetric theories and present arguments to support that the local supersymmetry continues to ensure the ANEC is satisfied even in general supersymmetric quantum field theories.}

\begin{document}
\maketitle

%%%%%%%%%%%%%%%%%%%%%%%%%%%%%%%%%%%%
\section{Introduction}
%%%%%%%%%%%%%%%%%%%%%%%%%%%%%%%%%%%%
Quantum field theory is a highly successful framework underlying much of modern physics. The most analyzed observables are scattering amplitudes and correlation functions of gauge invariant local operators. However, these are but a subset of a class of much more general observables that include \textit{non-local} operators. Their role has particularly been emphasized recently such as in the context of generalized symmetries and defects \cite{Gaiotto:2014kfa,Cordova:2022ruw}, the study of light-ray operators in conformal field theory \cite{Hofman:2008ar,Kravchuk:2018htv} to name a few. A unifying theme underlying these ideas and much of quantum field theory are symmetry principles that help us better understand the physics involved. These ideas have also unveiled and led to connections with diverse fields ranging from category theory to quantum information theory. Light ray operators, in particular, serve as powerful probes of Lorentzian field theory \cite{Hofman:2008ar,Faulkner:2016mzt,Hartman:2016lgu,Casini:2017roe,Kravchuk:2018htv,Meltzer:2018tnm,Cordova:2017dhq,Balakrishnan:2019gxl,Kologlu:2019mfz,Manenti:2019kbl,Belin:2020lsr,Chang:2020qpj,Korchemsky:2021htm,Caron-Huot:2022eqs,Belin:2026wkc}, play a central role in conformal collider physics \cite{Hofman:2008ar,Afkhami-Jeddi:2018own}, the Regge limit \cite{Kravchuk:2018htv}, the CFT inversion formula \cite{Caron-Huot:2017vep,Simmons-Duffin:2017nub,Kravchuk:2018htv}, holography \cite{Belin:2019mnx}, the realization of asymptotic symmetries \cite{Cordova:2018ygx,Besken:2020snx}, leading to predictions and connections to particle physics experiments \cite{Dixon:2019uzg,Komiske:2022enw,Lee:2022uwt,Moult:2025nhu} and have even been discussed in the context of simulating on a  quantum computer \cite{Lee:2024jnt}. They have also been central in recent proofs of monotonicity theorems for renormalization group flows \cite{Hartman:2023qdn,Hartman:2023ccw,Hartman:2024xkw}. Overall, there has been a lot of progress in their study and applications and they will form the main subject of this work.
\subsection{Summary}
Before we get into the details of our construction, we sketch our main idea and summarize our results.
In this paper, we focus on a set of universal light-ray operators and their algebra. We work with supersymmetric conformal field theories and construct light-ray operators using the stress tensor and supersymmetry current. One of these is the averaged null energy operator,
\begin{align}\label{ANECoperator}
    \mathcal{E}(x_{\perp})=\int_{-\infty}^{\infty}dx^- T_{--}(x^-,x^+=0,x_{\perp}).
\end{align}
The statement of its positivity in any state $|\Psi\rangle$ in the Hilbert space $\mathcal{H}$ of the field theory,
\begin{align}\label{ANEC}
    \langle \Psi|\mathcal{E}(x_{\perp})|\Psi\rangle\ge 0,
\end{align}
is known as the averaged null energy condition (ANEC). It was used by Hofman and Maldacena to derive bounds on OPE coefficients in conformal field theories \cite{Hofman:2008ar}. It was proven by Faulkner, Leigh, Parrikar, and Wang \cite{Faulkner:2016mzt} using information theoretic ideas such as the monotonicity of relative entropy and the related monotonicity under inclusions of the modular Hamiltonian. Their argument holds for generic interacting unitary Lorentz invariant quantum field theories. In \cite{Hartman:2016lgu}, it was also proved using microcausality, unitarity, Lorentz invariance, a twist gap and the lightcone operator product expansion. They also generalized the construction to positivity constraints on light-ray operators involving higher spin operators. 

In super-conformal field theories, the stress tensor is a part of a multiplet that includes the current that generates supersymmetry \cite{Dumitrescu:2011iu}, $\mathcal{J}_{\mu\alpha}^{I}$. $\mu$, $\alpha$, $I$ respectively denote vector, spinor and $R-$symmetry indices. We stick to $d=3$ and $d=4$ for concreteness in this paper but it should be possible to generalize most of what is to follow to higher dimensions. Central to our construction is the fermionic supersymmetric light-ray operator that we define,
\begin{align}
    q^{I}(x_{\perp})=\int_{-\infty}^{\infty}dx^-~\mathcal{J}^I_{-\downarrow}(x^-,x^+=0,x_{\perp}),
\end{align}
where $\downarrow$ denotes a particular spinor-component. We then constrain the anti-commutator of this operator with itself using microcausality and unitarity similar to the process in \cite{Cordova:2018ygx}, finding,
\begin{align}\label{qqgivesANEC}
    \{q^{I}(x_{1\perp}),q^\dagger_{J}(x_{2\perp})\}=2\delta^I_J\delta^{d-2}(x_{1\perp}-x_{2\perp})\mathcal{E}(x_{2\perp}).
\end{align}
Thus, $q^{I}(x_{\perp})$ is a local square-root of the averaged null energy operator. In general, we expect fermionic light-ray operators located at different transverse points in a null hypersurface to anti-commute, since any two distinct points on different null rays are space-like separated. However, this still allows for infinitely many contact terms (involving derivatives of delta functions). We then use conformal symmetry (twist counting) and unitarity to set all but one term to zero, resulting in \eqref{qqgivesANEC}. By appropriately smearing \eqref{qqgivesANEC} in the transverse coordinates, we then show that the ANEC follows from this ``light-ray supersymmetry", similar to how the positivity of the Hamiltonian follows from the usual supersymmetry algebra, providing an extremely simple derivation in super-conformal field theories. 

Further, we consider the bosonic light-ray operators, $\mathcal{K}(x_{\perp})$ and $\mathcal{N}_i(x_{\perp})$, also made out of the stress tensor and constructed by C{\'o}rdova and Shao in \cite{Cordova:2018ygx}. They showed that smeared light-ray operators formed using $\mathcal{E},\mathcal{K}$ and $\mathcal{N}_i$ obey the generalized $\mathfrak{bms}$ algebra. The $\mathfrak{bms}$ algebra is usually realized as the symmetry algebra of asymptotically flat space-times at null infinity \cite{Barnich:2009se,Strominger:2017zoo,Raclariu:2021zjz}. In a conformal field theory, one can perform a conformal transformation that maps say, future null infinity $x^+=\infty$, to a finite $x^+=0$ null hypersurface which is where we define our light-ray operators just like in \cite{Hofman:2008ar,Cordova:2018ygx}. They realize the supertranslations $\mathcal{T}(f)$ and superrotations $\mathcal{R}(Y^i)$ using these smeared light-ray operators, providing a purely field theory derivation of the $\mathfrak{bms}$ algebra. Given our supersymmetric light-ray operator $q^I(x_{\perp})$ is a square-root of $\mathcal{E}(x_{\perp})$ which when smeared gives the supertranslations $\mathcal{T}(f)$, it is natural to include it in the algebra. We constrain and compute its commutators with the super-rotations and show that we obtain a supersymmetric $\mathfrak{bms}$ algebra \cite{Banerjee:2015kcx,Barnich:2015sca,Banerjee:2016nio,Fuentealba:2020aax,Fotopoulos:2020bqj}. We do so for $\mathcal{N}-$extended super-conformal field theories in both $d=3$ and $d=4$ explicitly. In \cite{Guo:2024qzv}, the authors constructed a $\mathcal{N}=1$ supersymmetric $\mathfrak{bms}_4$ algebra in the free massless Wess-Zumino theory. In this paper, we perform an abstract analysis valid for any (strongly coupled) interacting SCFT. For theories with extended supersymmetry, we also construct a light-ray operator out of the $R-$symmetry current (which is part of the stress tensor multiplet) and show that in $d=3$, there is an additional infinite tower of symmetry generators that are added to the $\mathfrak{bms}_3$ algebra. However, in $d=4$, we find that the algebra does not close if we include it to the regular supersymmetric $\mathfrak{bms}_4$ algebra. In the context of $\mathcal{N}=8$ supergravity in four dimensional flat space, such an observation was made previously \cite{Banerjee:2022lnz}. Our result here provides a field theoretic derivation and a generalization to any supersymmetry. However, we show that this infinite dimensional $R-$symmetry tower can be included to a more general version of the supersymmetric $\mathfrak{bms}_4$ algebra where we relax the holomorphicity assumption on the smeared super-supersymmetry generators.

Let us now summarize our main result.

\textit{``We define a field theoretic realization of the $\mathcal{N}-$extended supersymmetric $\mathfrak{bms}$ algebra using smeared light-ray operators built out of the stress tensor, supersymmetry current and $R-$symmetry current. The central ingredient in this construction is the ``square root" of the averaged null energy operator which then ensures its positivity in any state".}

\subsection{Outline and Discussion}
In section \ref{sec:3d}, we discuss the construction of the supersymmetric $\mathfrak{bms}_3$ algebra. We begin with a review of C{\'o}rdova and Shao's construction specialized to $d=3$. We compactify the transverse space and obtain the commutation relations of the modes of the light-ray operators. We then discuss in full detail our light-ray supersymmetry construction for $\mathcal{N}=1$ super-conformal field theories. We constrain the various (anti-)commutators using microcausality, unitarity and closure, finding the generalized $\mathcal{N}=1$ supersymmetric $\mathfrak{bms}_3$ algebra. We then generalize to theories with extended supersymmetry, providing a field theoretic interpretation of its origin. Higher supersymmetry also entails an $R-$symmetry current using which we construct light-ray operators and derive the resulting larger algebra. In section \ref{sec:4d}, we discuss the construction in $d=4$, which brings about additional moving parts since the transverse space is two-dimensional. We construct the various light-ray operators and their (anti-)commutators, showing that they obey a generalized $\mathcal{N}-$extended supersymmetric $\mathfrak{bms}_4$ algebra. However, in contrast to $d=3$, we find that for any $\mathcal{N}$, the inclusion of the smeared $R-$symmetry light-ray operators to the usual supersymmetric $\mathfrak{bms}_4$ algebra leads to the algebra not closing, thus leaving only the global $R-$symmetry generator as part of it. However, we show that it is a consistent part of a generalized $\mathcal{N}-$extended supersymmetric $\mathfrak{bms}_4$ algebra.

One expects the general construction to generalize to higher dimensions but it is nevertheless interesting and important to explicitly prove it which we leave for a future work. Another interesting point is the effect of central extensions to the supersymmetry algebra. Given the general results of \cite{Ferrara:1997tx,Dumitrescu:2011iu}, it would be interesting to understand if there is some generalized construction of our supersymmetric light-ray algebra that is sensitive to central charge.

Another interesting direction is the following: Given the fact that our supersymmetric light-ray operator is constructed out of the supersymmetry current, which in holographic theories is dual to the gravitino, it would be interesting to understand the connection between soft theorems for supersymmetric theories on an AdS$_4$ and AdS$_5$ backgrounds and the present construction. The light-ray algebra we have constructed does not close under the action of the super-conformal algebra similar to its bosonic counterpart \cite{Cordova:2018ygx}. Strominger and Wei then showed that, in $d=3$, its completion leads beautifully to the $\Lambda-$deformed $\mathcal{L}_{\Lambda}w_{1+\infty}$ algebra \cite{Strominger:2026yrh}. There has been a lot of recent work on the connection between light-ray operators and infinite dimensional algebras in conformal field theory \cite{Sheta:2025oep,Himwich:2025ekg,Strominger:2026yrh,Himwich:2026exq,Heuveline:2026nxq,Zhu:2026ova,Goodenbour:2026wkp}. The present paper lays out the foundation for the supersymmetric completion and it would be very interesting to determine it.

A related question is the interpretation of how, if at all, the $\mathfrak{bms}$ and supersymmetric $\mathfrak{bms}$ algebras are realized in de-Sitter spacetime. All our constructions are intrinsically Lorentzian involving integrals over null-lines and so on. Thus, they do not generalize to Euclidean conformal field theories that are expected to holographically describe de-Sitter spacetime. It would be fascinating to explore whether one can still import some of these ideas over to this context.

Finally, coming back to the beginning of our discussion, we prove the averaged null energy condition in section \ref{sec:ANEC} as a simple consequence of the existence of the light-ray supersymmetry algebra. Our proof is for super-conformal field theories but the light-ray supersymmetry algebra at surface level appears to generalize to non-conformal supersymmetric quantum field theories as we discuss in appendix \ref{app:QFT}. We leave a more detailed analysis for the future. Making a connection to modular Hamiltonians following \cite{Faulkner:2016mzt} could also help establish a connection to information theoretic ideas and provide perspective and lead to a generalization of our results to supersymmetric quantum field theories. 

We also have a few appendices to supplement the material in the main-text. In appendix \ref{app:Notation}, we outline our notation and conventions in both three and four dimensional Minkowski spacetime. In appendix \ref{app:FreeTheory}, we verify our light-ray supersymmetry construction in the free massless Wess-Zumino theories in $d=3,4$. In appendix \ref{app:QFT}, we discuss a possible extension to generic supersymmetric quantum field theories. In appendix \ref{app:PoincareFromBMS}, we explicitly construct the null hyper-surface preserving conformal sub-group from the $\mathfrak{bms}$ algebra. We provide an alternate derivation of our construction in appendix \ref{app:commutators}. Finally, we discuss the cosmological constant deformation of the supersymmetric $\mathfrak{bms}_3$ algebras in appendix \ref{app:bmsToVirasoro}.

\section{Three dimensions}\label{sec:3d}
We begin in three dimensional Minkowski spacetime, $\mathbb{R}^{2,1}$, using light-cone coordinates. Our notation and conventions are given in appendix \ref{app:Notation}. We specialize C{\'o}rdova and Shao's construction of light-ray operators that obey the generalized $\mathfrak{bms}$ algebra \cite{Cordova:2018ygx} to $d=3$. We consider the stress tensor $T_{\mu\nu}(x)$ using which we define the bosonic light ray operators of interest. 
\begin{align}\label{EKN3d}
    &\mathcal{E}(y)=\int_{-\infty}^{\infty}dx^{-}T_{--}(x^{-},x^{+}=0,y),\notag\\
    &\mathcal{K}(y)=\int_{-\infty}^{\infty}dx^{-}x^{-}T_{--}(x^{-},x^{+}=0,y),\notag\\
    &\mathcal{N}_y(y)=\int_{-\infty}^{\infty}dx^{-}T_{-y}(x^{-},x^{+}=0,y).
\end{align} 
The first of these is the familiar averaged null energy operator. If we integrate these operators over the $y$ coordinate, we obtain the Poincare generators $P_{-},J_{tx}$ and $P_{y}$ respectively. We label these operators by their twist $\tau=\Delta-m$ where $m$ is the eigenvalue under the boost operator $J_{tx}$ and $\Delta$ is the scaling dimension. We work with the conventions of \cite{Cordova:2018ygx} where $x^{-}$ has boost weight $-1$, $\mathcal{E}(y)$ has $(\Delta,m,\tau)=(2,1,1)$. $\mathcal{K}(y)$ has $(\Delta,m,\tau)=(1,0,1)$ and $\mathcal{N}_y(y)$ has $(\Delta,m,\tau)=(2,0,2)$. These light-ray operators obey the following commutation relations.
\begin{align}\label{EKNcommutator3d}
\boxed{
\begin{aligned}
[\mathcal{E}(y_1),\mathcal{E}(y_2)]
&=0,
\\[4pt]
[\mathcal{K}(y_1),\mathcal{K}(y_2)]
&=0,
\\[4pt]
[\mathcal{K}(y_1),\mathcal{E}(y_2)]
&=-i\delta(y_1-y_2)\mathcal{E}(y_2),
\\[4pt]
[\mathcal{N}_{y}(y_1),\mathcal{E}(y_2)]
&=-i\delta(y_1-y_2)\partial_{y_2}\mathcal{E}(y_2)
+i\partial_{y_1}\delta(y_1-y_2)\mathcal{E}(y_2),
\\[4pt]
[\mathcal{N}_{y}(y_1),\mathcal{K}(y_2)]
&=-i\delta(y_1-y_2)\partial_{y_2}\mathcal{K}(y_2)
+i\partial_{y_1}\delta(y_1-y_2)\mathcal{K}(y_2),
\\[4pt]
[\mathcal{N}_{y}(y_1),\mathcal{N}_{y}(y_2)]
&=-i\delta(y_1-y_2)\partial_{y_2}\mathcal{N}_{y}(y_2)
+2i\partial_{y_1}\delta(y_1-y_2)\mathcal{N}_{y}(y_2).
\end{aligned}
}
\end{align}

We now define the smeared operators,
\begin{align}\label{TandRsmearingy}
    &\mathcal{T}(f)=\int_{-\infty}^{\infty}dy~f(y)~\mathcal{E}(y),\notag\\
    &\mathcal{R}(Y)=\bigg(\int_{-\infty}^{\infty} dy Y(y)\mathcal{N}_y(y)+\int_{-\infty}^{\infty}dy~\partial_y Y(y)~\mathcal{K}(y)\bigg).
\end{align}
These quantities obey the commutation relations,
\begin{align}\label{BMS3dyvariables}
    &[\mathcal{T}(f_1),\mathcal{T}(f_2)]=0,\notag\\
    &[\mathcal{T}(f),\mathcal{R}(Y)]=i\mathcal{T}(g),\notag\\
    &[\mathcal{R}(Y_1),\mathcal{R}(Y_2)]=i\mathcal{R}(Y_3),
\end{align}
where $g=(\partial_y Y)f-Y\partial_y f$ and $Y_3=Y_1\partial_y Y_2-(\partial_y Y_1)Y_2$. The commutation relations \eqref{BMS3dyvariables} define exactly the three dimensional generalized $\mathfrak{bms}$ algebra with $\mathcal{T}$ representing super-translations and $\mathcal{R}$ representing the super-rotations. To write it in a more familiar form, consider the compactification $y=\tan\big(\frac{\varphi}{2}\big)$ with $-\pi< \varphi< \pi$. We can transform the operators as \cite{Strominger:2026yrh} :
\begin{align}
    &\mathcal{E}(\phi)=\bigg(\frac{dy}{d\varphi}\bigg)^{2}\mathcal{E}(y)=\bigg(\frac{\sec^2\big(\frac{\varphi}{2}\big)}{2}\bigg)^{2}\mathcal{E}(y),\notag\\
    &\mathcal{K}(\varphi)=\bigg(\frac{dy}{d\varphi}\bigg)\mathcal{K}(y)=\bigg(\frac{\sec^2\big(\frac{\varphi}{2}\big)}{2}\bigg)\mathcal{K}(y),\notag\\
    &\mathcal{N}_y(y)=\mathcal{N}_\varphi(\varphi)\bigg(\frac{d\varphi}{dy}\bigg)^2+\mathcal{K}(\varphi)\bigg(\frac{d^2\varphi}{dy^2}\bigg)=\mathcal{N}_{\varphi}(\varphi)\bigg(\frac{2}{\sec^2\big(\frac{\varphi}{2}\big)}\bigg)^2+\mathcal{K}(\varphi)\bigg(\frac{-2\sin(\varphi)}{\sec^2\big(\frac{\varphi}{2}\big)}\bigg).
\end{align}
One can then obtain the algebra of the $\mathcal{E}(\varphi),\mathcal{K}(\varphi),\mathcal{N}_{\varphi}(\varphi)$ operators which takes the same form as when written in the $y$ variables \eqref{EKNcommutator3d}. For example,
\begin{align}
    [\mathcal{K}(\varphi_1),\mathcal{E}(\varphi_2)]&=\bigg(\frac{\sec^2\big(\frac{\varphi_1}{2}\big)}{2}\bigg)\bigg(\frac{\sec^2\big(\frac{\varphi_2}{2}\big)}{2}\bigg)^2[\mathcal{K}(y_1),\mathcal{E}(y_2)]\notag\\
    &=-i\delta\bigg(\tan\bigg(\frac{\varphi_1}{2}\bigg)-\tan\bigg(\frac{\varphi_2}{2}\bigg)\bigg)\bigg(\frac{\sec^2\big(\frac{\varphi_1}{2}\big)}{2}\bigg)\bigg(\frac{\sec^2\big(\frac{\varphi_2}{2}\big)}{2}\bigg)^2\mathcal{E}(y_2)\notag\\
    &=-i\bigg(\frac{\sec^2\big(\frac{\varphi_1}{2}\big)}{2}\bigg)\sum_{k=-\infty}^{\infty}\frac{2\delta(\varphi_1-\varphi_2-2\pi k)}{\sec^2\big(\frac{\phi_2+2\pi k}{2}\big)}\mathcal{E}(\varphi_2)\notag\\
    &=-i\delta(\varphi_1-\varphi_2)\mathcal{E}(\varphi_2),
\end{align}
where we used the fact that only the $k=0$ contribution from the Dirac comb function contributes due to the range $-\pi<\phi_1,\phi_2<\pi$. One can similarly show that all other commutators take the same form as \eqref{EKNcommutator3d} with $y$ replaced by $\varphi$. We now define operators smeared over the angular variable $\varphi$ exactly as in \eqref{TandRsmearingy}\footnote{The functions appearing in \eqref{TandRsmearingphi} are related to those in \eqref{TandRsmearingy} by $f(\varphi)=2\cos^2\big(\frac{\varphi}{2}\big)f(y=\tan(\frac{\varphi}{2}))$ and $Y_\varphi(\varphi)=2\cos^2\big(\frac{\varphi}{2}\big)Y_y(y=\tan(\frac{\varphi}{2}))$.}.
\begin{align}\label{TandRsmearingphi}
    &\mathcal{T}(f)=\int_{-\pi}^{\pi}d\varphi~f(\varphi)~\mathcal{E}(\varphi),\notag\\
    &\mathcal{R}(Y)=\bigg(\int_{-\pi}^{\pi}d\varphi Y_{\varphi}(\varphi)\mathcal{N}_y(\varphi)+\int_{-\pi}^{\pi}d\varphi~\partial_\varphi Y_{\varphi}(\varphi)~\mathcal{K}(\varphi)\bigg).
\end{align}
These quantities obey commutation relations identical to those in \eqref{BMS3dyvariables}.
\begin{align}\label{BMS3dphivariables}
    &[\mathcal{T}(f_1),\mathcal{T}(f_2)]=0,\notag\\
    &[\mathcal{T}(f),\mathcal{R}(Y)]=i\mathcal{T}(g),\notag\\
    &[\mathcal{R}(Y_1),\mathcal{R}(Y_2)]=i\mathcal{R}(Y_3),
\end{align}
We then define the Fourier modes,
\begin{align}\label{3dMandLmodes}
    M_n=\mathcal{T}(e^{in\varphi}),~L_n=\mathcal{R}(e^{i n\varphi}),~~m,n\in\mathbb{Z}.
\end{align}
We find the familiar $\mathfrak{bms}_3$ mode algebra \cite{Barnich:2010eb},
\begin{align}\label{bosonic3dBMS}
    &[M_n,M_m]=0,\notag\\
    &[L_m,M_n]=(m-n)M_{m+n},\notag\\
    &[L_m,L_n]=(m-n)L_{m+n}.
\end{align}

We now proceed to theories with supersymmetry where we will define additional light-ray operators that together with \eqref{BMS3dphivariables}, will result in a supersymmetric $\mathfrak{bms}$ algebra.
\subsection{$\mathcal{N}=1$ supersymmetry}
The defining relation of $\mathcal{N}=1$ supersymmetry is the following anti-commutation relation.
\begin{align}
    \{Q_{a},Q_{b}\}=P_{ab}.
\end{align}
We focus on the component $Q_1=Q_{\downarrow}$ which satisfies,
\begin{align}
    \{Q_{\downarrow},Q_{\downarrow}\}=2P_-.
\end{align}
Please see appendix \ref{app:Notation} for more details on the notation. We consider the supersymmetry current $\mathcal{J}_{\mu a}(x)$ which gives rise to the conserved charges $Q_a$ as follows:
\begin{align}
    Q_a=\int_{-\infty}^{\infty}dy\int_{-\infty}^{\infty}dx^- \mathcal{J}_{-\downarrow}(x^-,x^+=0,y).
\end{align}
Motivated by the relationship between $P_-$ and the averaged null energy operator $\mathcal{E}$ we define the following fermionic light-ray operator.
\begin{align}\label{qNeq13d}
    q(y)=\int_{-\infty}^{\infty}dx^-\mathcal{J}_{-\downarrow}(x^-,x^+=0,y).
\end{align}
The supersymmetry current has scaling dimension $\frac{5}{2}$ and boost weight $\frac{3}{2}$. Thus, we see that $q(y)$ has $(\Delta,m,\tau)=(\frac{3}{2},\frac{1}{2},1)$. If we integrate this quantity with respect to $y$, we obtain the supercharge $Q_\downarrow$. Our goal now is to derive the supersymmetric completion of the algebra \eqref{bosonic3dBMS}. This requires us to determine the various anti-commutation relations involving the fermionic light-ray operator \eqref{qNeq13d} and the bosonic ones defined earlier \eqref{EKN3d}. Our assumptions are the same as \cite{Cordova:2018ygx} with the inclusion of fermionic statistics.
\begin{itemize}
    \item Microcausality: space-like separated operators commute if they are bosonic and anti-commute if they are fermionic. If one is bosonic and the other is fermionic, they commute.
    \item Unitarity. All local operators transform in unitary representations of the double cover of $SO(3,2)$, which is the three dimensional conformal group. In particular, we use the unitarity of the super-conformal field theory to constrain the twists $\tau$ of local operators to be greater than or equal to $1$ \cite{Minwalla:1997ka}.
   \item Poincare transformations: $P_{-},J_{tx},P_{y}$ implement the appropriate Poincare transformations.
   \item Closure: Another technical and essential point is that the commutator of parallel light-ray operators that are integrals of conserved currents, can also be expressed as integrals of local operators. We justify this assumption in appendix \ref{app:commutators}.
\end{itemize} 
With the operators at hand and the assumptions laid out, we proceed to the computation of the (anti-)commutators.
\subsubsection*{$\{q,q\}$}
We have,
\begin{align}
    \{q(y_1),q(y_2)\}=\frac{1}{4}\int_{-\infty}^{\infty}dx_1^{-}\int_{-\infty}^{\infty}dx_2^{-}\{\mathcal{J}_{\downarrow\downarrow\downarrow}(x_1^{-},0,y_1),\mathcal{J}_{\downarrow\downarrow\downarrow}(x_2^{-},0,y_2)\}.
\end{align}
The two supersymmetry currents are spacelike separated since they are both at $x^+=0$. Thus, microcausality implies that this anti-commutator vanishes unless $y_1=y_2$. A general ansatz is,
\begin{align}\label{QQeq1}
    \{q(y_1),q(y_2)\}=\delta(y_1-y_2)A_0(y_1)+\sum_{n=1}^{\infty}\partial_{y_1}^{n}\delta(y_1-y_2)A_n(y_1).
\end{align}
The total twist on the LHS is $1+1=2$ which implies that the $A_n$ have twist $\tau=1-n$. Closure implies that the $A_n$ themselves can be written as a null integral of some local operator $\phi_n$ as $A_n(y)=\int_{-\infty}^{\infty}dx^- \phi_n(x^-,0,y)$. This tells us that $\phi_n$ has twist $\tau_{\phi_n}=1-n$ since $dx^-$ does not contribute to the twist. Therefore, unitarity implies that $A_n,n>0$ should vanish\footnote{Unitarity bounds for scalars state that their twist $\tau\ge \frac{1}{2}$ in three dimensions. However, since $q(y)$ has boost eigenvalue $m=\frac{1}{2}$, its anti-commutator has $m=1$ and thus scalars cannot contribute. We could also allow for the existence of a tower of higher spin currents $J_s^{\mu_1\cdots \mu_s}$ (in which case we are dealing with a free theory). All these operators have $\tau=1$. However, by dimensional analysis, they cannot contribute to this anti-commutator since they have scaling dimension $s+1$.}. Thus we are left with just the first term on the RHS. To determine $A_0$, we integrate both sides of the equation \eqref{QQeq1} with respect to $y_1$. This yields,
\begin{align}
    \{Q_{\downarrow},q(y)\}=A_0(y).
\end{align}
$\mathcal{N}=1$ Supersymmetry tells us that,
\begin{align}
     \{Q_{\downarrow},q(y)\}=\frac{1}{2}\int_{-\infty}^{\infty}dx^{-}\{Q_{\downarrow},\mathcal{J}_{\downarrow\downarrow\downarrow}(x^{-},0,y)\}=2\int_{-\infty}^{\infty}dx^{-}T_{--}(x^{-},0,y)=2\mathcal{E}(y).
\end{align}
As a consistency check, we integrate over the remaining $y$ coordinate which yields,
\begin{align}
    \{Q_{\downarrow},Q_{\downarrow}\}=2 P_{-},
\end{align}
which is the correct supersymmetry algebra relation. Thus, we have established,
\begin{align}\label{qqANEC3d}
\boxed{
\{q(y_1),q(y_2)\}=2\delta(y_1-y_2)\mathcal{E}(y_1).
}
\end{align}
$q(y)$ can thus be interpreted as a local fermionic square-root of the averaged null energy operator. We will study the implications of this equation for the ANEC in section \ref{sec:ANEC}.
\subsubsection*{$[\mathcal{E},q]$}
By unitarity and microcausality we find,
\begin{align}
    [\mathcal{E}(y_1),q(y_2)]=\delta(y_1-y_2)A_0(y_2).
\end{align}
Integrating both sides of this equation with respect to $y_1$ results in,
\begin{align}
    [P_-,q(y)]=A_0(y)=0,
\end{align}
since $q(y)$ is translation invariant along the $x^-$ direction. Therefore,
\begin{align}
    [\mathcal{E}(y_1),q(y_2)]=0.
\end{align}
\subsubsection*{$[\mathcal{K},q]$}
Let us now compute the commutator of $q$ with $\mathcal{K}(y)$. Using microcausality and unitarity we can write,
\begin{align}
    [\mathcal{K}(y_1),q(y_2)]=\delta(y_1-y_2)A(y_1).
\end{align}
$q$ has twist $1$ and $\mathcal{K}$ has twist $1$ and by the same unitarity argument as before, we have ruled out terms in the ansatz involving derivatives of delta functions. Integrating both sides of the above equation with respect to $y_2$ results in,
\begin{align}
    [J_{tx},q(y)]=A(y).
\end{align}
From its definition, we see that $q(y)$ has (boost) eigen value $\frac{1}{2}$, thus fixing $A(y)=-\frac{i}{2}q(y)$. We have thus found our second commutator:
\begin{align}
\boxed{
[\mathcal{K}(y_1),q(y_2)]=-\frac{i}{2}\delta(y_1-y_2)q(y_1).
}
\end{align}
\subsection*{$[\mathcal{N}_y,q]$}
Next, we consider the commutator with $\mathcal{N}_y(y)$. We have,
\begin{align}
    [\mathcal{N}_y(y_1),q(y_2)]=\delta(y_1-y_2)A_0(y_2)+\partial_{y_1}\delta(y_1-y_2)~A_1(y_2).
\end{align}
$\mathcal{N}_y$ has twist $2$ which implies that $A_0$ has twist $2$ and $A_1$ has twist $1$. Higher derivative terms are inconsistent with unitarity.
Integrating both sides with respect to $y_1$ results in,
\begin{align}
    [P_y,q(y)]=A_0(y).
\end{align}
By the Poincare algebra we know that this equals $-i\frac{\partial}{\partial y}q(y)$ which gives an equation,
\begin{align}
    A_0(y)=-i\partial_y q(y).
\end{align}
To determine $A_1$, we integrate with respect to $y_2$ which results in,
\begin{align}
    [\mathcal{N}_y(y_1),Q_{\downarrow}]=A_0(y_1)+\partial_{y_1}A_1(y_1).
\end{align}
From their definitions we have\footnote{Naively, the commutator $[Q_{\downarrow},T_{-y}(x^-,0,y)]$ can contain a $\partial_y \mathcal{J}_{-\downarrow}$ but a careful analysis of the supersymmetry transformations shows that this term does not occur.},
\begin{align}
    &[\mathcal{N}_y(y_1),Q_{\downarrow}]=\int_{-\infty}^{\infty}dx^{-} [T_{-y}(x^-,0,y),Q_{\downarrow}]=-\frac{1}{2}\int_{-\infty}^{\infty}dx^{-}[Q_{\downarrow},T_{\downarrow\downarrow \downarrow \uparrow}(x^-,0,y)]\notag\\
    &=\text{boundary term which vanishes}=0.
\end{align}
Therefore, we find,
\begin{align}
    0=-i\partial_y q(y)+\partial_y A_1(y)\implies A_1(y)=iq(y).
\end{align}
Thus we have found the relation,
\begin{align}
    \boxed{[\mathcal{N}_y(y_1),q(y_2)]=-i\delta(y_1-y_2)\partial_{y_2}q(y_2)+i\partial_{y_1}\delta(y_1-y_2)q(y_2).}
\end{align}
To summarize, we have found that the bosonic algebra obeyed by $\mathcal{E},\mathcal{K}$ and $\mathcal{N}_y$ \cite{Cordova:2018ygx} should be supplemented by the following (anti-)commutators involving the fermionic light-ray operator $q$ to obtain its $d=3$, $\mathcal{N}=1$ supersymmetric counterpart.
\begin{align}\label{Neq13dQcommutators}
\boxed{
\begin{aligned}
\{q(y_1),q(y_2)\}
&=
2\delta(y_1-y_2)\mathcal{E}(y_2),
\\[4pt]
[\mathcal{E}(y_1),q(y_2)]&=0
\\[4pt]
[\mathcal{N}_y(y_1),q(y_2)]
&=
-i\delta(y_1-y_2)\partial_{y_2}q(y_2)
+i\partial_{y_1}\delta(y_1-y_2)
q(y_2),
\\[4pt]
[\mathcal{K}(y_1),q(y_2)]
&=
-\frac{i}{2}\delta(y_1-y_2)
q(y_2).
\end{aligned}
}
\end{align}
We now define the smeared operator,
\begin{align}
    \mathcal{Q}(\epsilon)=\int_{-\infty}^{\infty}dy~\epsilon(y)q(y).
\end{align}
Using the relations \eqref{Neq13dQcommutators}, we can now compute the (anti-)commutators of this operator and the ones defined in \eqref{TandRsmearingy}. \begin{align}
    \{\mathcal{Q}(\epsilon_1),\mathcal{Q}(\epsilon_2)\}&=\int_{-\infty}^{\infty}dy_1\int_{-\infty}^{\infty}dy_2 \epsilon_1(y_1)\epsilon_2(y_2)\{q(y_1),q(y_2)\}=2\int_{-\infty}^{\infty}dy~(\epsilon_1(y)\epsilon_2(y))\mathcal{E}(y)\notag\\
    &=2\mathcal{T}(\epsilon_1\epsilon_2).
\end{align}
It is also easy trivial to check that,
\begin{align}
    [\mathcal{Q}(\epsilon),\mathcal{T}(f)]=0.
\end{align}
Finally,
\begin{align}
    [\mathcal{Q}(\epsilon),\mathcal{R}(Y)]&=\int_{-\infty}^{\infty}dy_1\int_{-\infty}^{\infty}dy_2\bigg(\epsilon(y_1)Y(y_2)[q(y_1),\mathcal{N}_y(y_2)]+\partial_{y_2}Y(y_2)\epsilon(y_1)[q(y_1),\mathcal{K}(y_2)]\bigg)\notag\\
    &=\int_{-\infty}^{\infty}dy_1\int_{-\infty}^{\infty}dy_2\bigg(\epsilon(y_1)Y(y_2)\big(i\delta(y_1-y_2)\partial_{1}q(y_1)-i\partial_2\delta(y_1-y_2)~q(y_1)\big)\notag\\
    &\qquad\qquad\qquad+\frac{i}{2}\partial_{y_2}Y(y_2)\epsilon(y_1)\delta(y_1-y_2)q(y_1)\bigg)\notag\\
    &=\int_{-\infty}^{\infty}dy_2\bigg(i\epsilon(y_2)Y(y_2)\partial_2 q(y_2)+i\epsilon(y_2)\partial_2Y(y_2)~q(y_2)+\frac{i}{2}\partial_{y_2}Y(y_2)\epsilon(y_1)q(y_1)\bigg)\notag\\
    &=\int_{-\infty}^{\infty}dy\bigg(-i\epsilon(y)\partial_y Y(y)-i\partial_y \epsilon(y)~Y(y)+i\epsilon(y) \partial_y Y(y)+\frac{i}{2}\partial_y Y(y)~\epsilon(y)\bigg)q(y)\notag\\
    &=-i \mathcal{Q}(\psi),
\end{align}
where $\psi(y)=Y(y)\partial_y\epsilon(y)-\frac{1}{2}\epsilon(y)\partial_y Y(y)$. Along with the bosonic commutators \eqref{BMS3dyvariables}, these (anti-)commutators define the generalized supersymmetric $\mathfrak{bms}_3$ algebra.

To bring this algebra to a more familiar form, we compactify the $y$ direction and introduce modes as in \eqref{3dMandLmodes} for the $q$ operator. We define the smeared operator\footnote{$q(\varphi)=\frac{dy}{d\varphi}q(y)$ is the transformation property under the map. The form of the (anti-)commutators \eqref{Neq13dQcommutators} is unchanged and obtained just by replacing $y\to\varphi$.},
\begin{align}\label{QNeq1smeared}
    \mathcal{Q}(\epsilon)=\int_{-\pi}^{\pi}d\varphi~\epsilon(\varphi)q(\varphi).
\end{align}
It is easy to check that we obtain an identical algebra,
\begin{align}\label{QTRNeq1Algebra1}
    &\{\mathcal{Q}(\epsilon_1),\mathcal{Q}(\epsilon_2)\}=2\mathcal{T}(\epsilon_3),\notag\\
    &[\mathcal{Q}(\epsilon),\mathcal{T}(f)]=0,\notag\\
    &[\mathcal{Q}(\epsilon_1),\mathcal{R}(Y)]=-i\mathcal{Q}(\psi),
\end{align}
where $\epsilon_3(\varphi)=\epsilon_1(\varphi)\epsilon_2(\varphi)$ and $\psi(\varphi)=Y_{\varphi}(\varphi)\partial_\varphi \epsilon(\varphi)-\frac{1}{2}\epsilon(\varphi)\partial_{\varphi}Y(\varphi)$. 
We now extract a particular Fourier mode of $\mathcal{Q}(\epsilon)$ via,
\begin{align}
    &F_r=\mathcal{Q}(e^{i r \varphi}), r\in\mathbb{Z}+\frac{1}{2}~(\text{Neveu Schwartz}),\notag\\
    &F_r=\mathcal{Q}(e^{i r \varphi}),r\in\mathbb{Z}~(\text{Ramond}).
\end{align}
The distinction between the anti-periodic and periodic choices will not matter in what follows. With the bosonic mode commutators \eqref{bosonic3dBMS} and the (anti-)commutation relations \eqref{QTRNeq1Algebra1}, we find,
\begin{align}\label{fermionicNeq13dBMS}
    &\{F_r,F_s\}=2 M_{r+s},[F_r,M_n]=0,[F_r,L_m]=(r-\frac{m}{2})F_{r+m}.
\end{align}
\begin{3dBMSNeq1}
Thus we find light-ray operators in $\mathcal{N}=1$ SCFT$_3$ which generate the supersymmetric $\mathfrak{bms}_3$ algebra:
\begin{align}\label{3dNeq1BMS}
   &[M_n,M_m]=0,\notag\\
    &[L_m,M_n]=(m-n)M_{m+n},\notag\\
    &[L_m,L_n]=(m-n)L_{m+n},\notag\\
     &\{F_r,F_s\}=2 M_{r+s},\notag\\
    &[F_r,M_n]=0,\notag\\
    &[F_r,L_m]=\bigg(r-\frac{m}{2}\bigg)F_{r+m}.
\end{align}
\end{3dBMSNeq1}
Therefore, any unitary $\mathcal{N}=1$ super-conformal field theory in three dimensions contains a universal $\mathfrak{bms}_3$ sub-algebra generated by smeared light-ray operators \eqref{TandRsmearingphi} and \eqref{QNeq1smeared}.

\subsection{Extended supersymmetry}
We proceed to the generalization to theories with extended supersymmetry. The novel ingredient is the presence of a $R-$symmetry current $J_{(R)\mu}$ which rotates the supersymmetry generators amongst themselves, keeping the momentum fixed. In three dimensions, the $R-$symmetry group is $SO(\mathcal{N})$. Thus, for $\mathcal{N}-$extended supersymmetry, there are a total of $2\mathcal{N}$ supersymmetry generators\footnote{If we include the special super-conformal generators, the total number goes to $4\mathcal{N}$. For $\mathcal{N}=8$, we get $32$ fermionic generators in the maximal supersymmetric case which is appropriate for the BLG theory.}. 
The defining relation of $\mathcal{N}-$extended supersymmetry is the following anti-commutation relation.
\begin{align}
    \{Q^{I}_{a},Q^{J}_{b}\}=\delta^{IJ}P_{ab}.
\end{align}
In particular, we focus on the $a=b=1=\downarrow$ components which obey,
\begin{align}
    \{Q_{\downarrow}^{I},Q_{\downarrow}^{J}\}=2\delta^{IJ}P_{-}.
\end{align}
These quantities are conserved charges constructed using the supersymmetry current $\mathcal{J}^I_{abc}$.
\begin{align}
    Q^I_{a}=\int d\Sigma^{ab}\mathcal{J}^I_{abc}=\frac{1}{2}\int_{-\infty}^{\infty}dy\int_{-\infty}^{\infty} dx^{-}~\mathcal{J}^I_{a\downarrow\downarrow}(x^{-},x^{+}=0,y).
\end{align}
We now define the following fermionic light-ray operator,
\begin{align}\label{QAnecNextended3d}
    q^I(y)=\frac{1}{2}\int_{-\infty}^{\infty}dx^{-}\mathcal{J}^I_{\downarrow \downarrow\downarrow}(x^{-},x^{+}=0,y).
\end{align}
 Integrating this quantity over the $y$ coordinate, we obtain the supersymmetry generators $Q^I_{\downarrow}$. In contrast to $\mathcal{N}=1$, we now consider the $R-$symmetry conserved current. Clearly, this current is in the adjoint representation of $SO(\mathcal{N})$ and we denote it as $J_{(R)}^{IJ}$, which is anti-symmetric in the $I,J$ indices. For $\mathcal{N}=2$ supersymmetry, the adjoint representation is the singlet and thus we can trade this object for $\frac{\epsilon^{IJ}}{2}J_{(R)}$ but for general $\mathcal{N}$, we work with $J^{IJ}_{(R)}$. This current, for general $\mathcal{N}-$extended supersymmetry, is part of the stress tensor supermultiplet. We define the $R-$charge,
 \begin{align}
     R^{IJ}=\int_{-\infty}^{\infty}dy\int_{-\infty}^{\infty}dx^-J_{(R)-}^{IJ}(x^{-},x^{+}=0,y).
 \end{align}
It rotates the supersymmetry generators amongst themselves,
\begin{align}
    [R^{IJ},Q^K_a]=\delta^{I K}Q_{a}^{J}-\delta^{JK}Q_{a}^{I}.
\end{align}
We now define the $R-$charge light-ray operator,
\begin{align}
    r^{IJ}(y)=\int_{-\infty}^{\infty}dx^-J_{(R)}^{IJ}(x^-,x^{+}=0,y).
\end{align}
This enlarges our set of light-ray operators to,
\begin{align}
    \{\mathcal{E},\mathcal{K},\mathcal{N}_{i},q^I,r^{IJ}\}.
\end{align}
Essentially, we construct a light-ray operator out of every conserved current that is part of the stress tensor multiplet.
Our task now is to compute the various (anti-)commutators between these light-ray operators and determine the algebra that they obey. The steps involved are quite analogous to those for $\mathcal{N}=1$. We make the same assumptions: Microcausality, unitarity, in particular the bound on twists, Poincare invariance and closure. Since the steps involved are quite analogous to the $\mathcal{N}=1$ construction, we do not present the details of the derivation. In addition to the usual bosonic sub-algebra, we find the following additional relations:
\begin{align}\label{Neq2qralgebray}
    &\{q^{I}(y_1),q^{J}(y_2)\}=2\delta^{IJ}\delta(y_1-y_2)\mathcal{E}(y_2)\notag\\
    &[\mathcal{E}(y_1),q^{I}(y_2)]=0,\notag\\
    &[\mathcal{K}(y_1),q^{I}(y_2)]=-\frac{i}{2}\delta(y_1-y_2)q^{I}(y_2),\notag\\
    &[\mathcal{N}_y(y_1),q^{I}(y_2)]=-i\delta(y_1-y_2)\partial_{y_2}q^{I}(y_2)+i\partial_{y_1}\delta(y_1-y_2)q^{I}(y_2),\notag\\
    &[r^{IJ}(y_1),q^K(y_2)]=\delta(y_1-y_2)\bigg(\delta^{IK}q^{J}(y_2)-\delta^{JK}q^{K}(y_2)\bigg),\notag\\
    &[\mathcal{N}_y(y_1),r^{IJ}(y_2)]=-i\delta(y_1-y_2)\partial_{y_2}r^{IJ}(y_2)+i\partial_{y_1}\delta(y_1-y_2)r^{IJ}(y_2)\notag\\
    &[r^{IJ}(y_1),r^{KL}(y_2)]=\delta(y_1-y_2)\bigg(\delta^{IK}r^{JL}-\delta^{IL}r^{JK}-\delta^{JK}r^{IL}+\delta^{JL}r^{IK}\bigg).
\end{align}
with the remaining commutators between $r^{IJ}$ and the operators vanishing. In particular, it commutes with $\mathcal{K}(y)$ which can be seen using the fact that it has zero boost weight. We can now transform these quantities from $\mathbb{R}$ to $S^1$ by the compactification map $y=\tan\big(\frac{\varphi}{2}\big)$ which results in the algebra \eqref{Neq2qralgebray} with $y$ replaced by $\varphi$. We then define the smeared operators,
\begin{align}
    &\mathcal{Q}^{I}(\epsilon)=\int_{-\pi}^{\pi}d\varphi~\epsilon(\varphi)q^{I}(\varphi),\notag\\
    &\rho^{IJ}(f)=\int_{-\pi}^{\pi}d\varphi~f(\varphi)r^{IJ}(\varphi).
\end{align}
Along with the smeared super-translation and super-rotation operators \eqref{TandRsmearingphi} and the commutation relations \eqref{Neq2qralgebray}, one finds the following non-zero (anti-)commutators:
\begin{align}
&\{\mathcal{Q}^{I}(\epsilon_1),\mathcal{Q}^{J}(\epsilon_2)\}=2\delta^{IJ}\mathcal{T}(\epsilon_1\epsilon_2),\notag\\&[\mathcal{Q}^{I}(\epsilon),\mathcal{T}(f)]=0,\notag\\
    &[\mathcal{Q}^{I}(\epsilon),\mathcal{R}(Y_{\varphi})]=-i\mathcal{Q}^{I}(\psi),\notag\\
    &[\mathcal{R}(Y),\rho^{IJ}(f)]=-i\rho^{IJ}(g),\notag\\
    &[\rho^{IJ}(f_1),\rho^{KL}(f_2)]=\delta^{Ik}\rho^{JL}(f_1 f_2)-\delta^{IL}\rho^{JK}(f_1f_2)-\delta^{JK}\rho^{IL}(f_1f_2)+\delta^{JL}\rho^{IK}(f_1f_2),
\end{align}
where $g=(\partial_{\varphi}Y)f-Y\partial_{\varphi}f$, $\psi=Y_{\varphi}\partial_{\varphi}\epsilon-\frac{1}{2}\epsilon\partial_{\varphi}Y_{\varphi}$. We then define the Fourier modes,
\begin{align}
    &F^I_r=\mathcal{Q}^I(e^{ir\varphi}),\notag\\
    &\rho^{IJ}_m=\rho^{IJ}(e^{im\varphi}),
\end{align}
where $r\in\mathbb{Z}$ or $r\in \mathbb{Z}+\frac{1}{2}$ depending on the periodicity conditions we wish to impose and $m\in\mathbb{Z}$. We can then go ahead and compute the (anti-)commutators of these modes using the general relations we have derived above.
\begin{3dBMSNextended}
Thus, we have determined light-ray operators in $\mathcal{N}-$extended SCFT$_3$ which generate the supersymmetric $\mathfrak{bms}_3$ algebra:
\begin{align}\label{3dsusybmsNextended}
   &[M_n,M_m]=0,\notag\\
    &[L_m,M_n]=(m-n)M_{m+n},\notag\\
    &[L_m,L_n]=(m-n)L_{m+n},\notag\\
     &\{F_r^I,F_s^J\}=2\delta^{IJ} M_{r+s},\notag\\
    &[F_r^I,M_n]=0,\notag\\
    &[F_r^I,L_m]=\bigg(r-\frac{m}{2}\bigg)F^I_{r+m},\notag\\
    &[\rho_m^{IJ},F_r^{K}]=\delta^{IK}F_{r+m}^{J}-\delta^{JK}F_{r+m}^{I},\notag\\
    &[L_m,\rho_n^{IJ}]=-n\rho_{m+n}^{IJ},\notag\\
    &[\rho_m^{IJ},\rho_n^{KL}]=\delta^{IK}\rho_{m+n}^{JL}-\delta^{IL}\rho_{n+m}^{JK}-\delta^{JK}\rho_{n+m}^{IL}+\delta^{JL}\rho_{n+m}^{IK}.
\end{align}
\end{3dBMSNextended}
Therefore, we have proved under reasonable assumptions that every $\mathcal{N}-$extended SCFT$_3$ contains a universal supersymmetric $\mathfrak{bms}_3$ sub-algebra.
\section{Four dimensions}\label{sec:4d}
We now proceed to four dimensional Minkowski spacetime. Our notation and conventions are provided in appendix \ref{app:Notation}. The main distinction from $d=3$ is the fact that the transverse space (representative of the celestial sphere or torus if we are in Klein signature when comparing with celestial holography) is two dimensional. This creates two copies of super-rotations (holomorphic and anti-holomorphic) and also leads to an extra term in the transformation of the supersymmetric light-ray operators under the action of the $\mathcal{N}_i$ generator to account for its transverse spin. Supertranslation modes are also labeled by two integers. The C{\'o}rdova-Shao operators in four dimensional Lorentzian spacetime take the form,
\begin{align}\label{EKN4d}
    &\mathcal{E}(z,\Bar{z})=\int_{-\infty}^{\infty}dx^{-}T_{--}(x^{-},x^{+}=0,z,\Bar{z}),\notag\\
    &\mathcal{K}(z,\Bar{z})=\int_{-\infty}^{\infty}dx^{-}x^{-}T_{--}(x^{-},x^{+}=0,z,\Bar{z}),\notag\\
    &\mathcal{N}_i(z,\Bar{z})=\int_{-\infty}^{\infty}dx^{-}T_{-i}(x^{-},x^{+}=0,z,\Bar{z}),
\end{align}
where $i$ runs over the transverse coordinates $(x^1,x^2)$ or equivalently $(z,\Bar{z})$. In $d=4$, a symmetric traceless conserved current has twist $\tau=\Delta-s=s+d-2-s=d-2=2$. Normalizing our boost generator such that $x^-$ has weight $-1$ then tells us that $\mathcal{E}$ has $(\Delta,s,\tau)=(3,1,2)$, $\mathcal{K}$ has $(\Delta,s,\tau)=(2,0,2)$ and $\mathcal{N}_i$ has $(\Delta,s,\tau)=(3,0,3)$. We then define the smeared operators,
\begin{align}
    &\mathcal{T}(f)=\int d^2 z~f(z,\Bar{z})\mathcal{E}(z,\Bar{z}),\notag\\
    &\mathcal{R}(Y)=\int d^2 z\bigg(Y^i(z,\Bar{z})\mathcal{N}_i(z,\Bar{z})+\frac{1}{2}\partial_i Y^i(z,\Bar{z})~\mathcal{K}(z,\Bar{z})\bigg).
\end{align}
The commutation relations are,
\begin{align}\label{4dbosonicfunctionbms}
    &[\mathcal{T}(f_1),\mathcal{T}(f_2)]=0,\notag\\
    &[\mathcal{T}(f),\mathcal{R}(Y)]=i\mathcal{T}(g),\notag\\
    &[\mathcal{R}(Y_1),\mathcal{R}(Y_2)]=i\mathcal{R}(Y_3),
\end{align}
where $g=\frac{1}{2}(\partial_i Y^i)f-Y^i\partial_i f$ and $Y_3^i=Y_1^j\partial_j Y_2^i-(\partial_j Y_1^i)Y_2^j$. We now define the modes of these operators via\footnote{The above commutation relations are when the transverse space is $\mathbb{R}^{2}$ so these are Laurent modes. However, while expressing them in terms of modes, we could 
choose to compactify this space to $S^2$, obtaining a discrete set of harmonics like we did in three dimensions. More rigorously, one can perform a stereographic projection from $\mathbb{R}^{2}$ to $S^2$ which will ultimately lead to the same results as we have seen in explicit detail in $d=3$.},
\begin{align}\label{4dbosonicmodes}
    &\qquad\qquad\qquad\qquad  M_{r,s}=i\mathcal{T}(z^r \Bar{z}^s),\notag\\
    &L_n=i\mathcal{R}(Y^i=-\delta^i_z z^{n+1}),~~\Bar{L}_n=i\mathcal{R}(Y^i=-\delta^i_{\Bar{z}}\Bar{z}^{n+1}).
\end{align}
It is then easy to verify that,
\begin{align}\label{4dbosonicbms}
    &\qquad\qquad\qquad\qquad\qquad\qquad[M_{r,s},M_{r',s'}]=0,\notag\\
    &[L_n,M_{r,s}]=\bigg(r-\frac{(n+1)}{2}\bigg)M_{r+n,s},~~[\Bar{L}_n,M_{r,s}]=\bigg(s-\frac{(n+1)}{2}\bigg)M_{r,s+n},\notag\\
    &[L_m,L_n]=(m-n)L_{m+n},\qquad[L_m,\Bar{L}_n]=0,\qquad[\Bar{L}_m,\Bar{L}_n]=(m-n)\Bar{L}_{m+n},
\end{align}
which is the familiar form of the usual $\mathfrak{bms}_4$ algebra \cite{Barnich:2009se}. Let us now generalize this construction to include supersymmetry. 
\subsection{The supersymmetric construction}
In $d=4$, $\mathcal{N}-$extended supersymmetric theories, the defining relation is,
\begin{align}
    \{Q_{\alpha}^{I},\tilde{Q}_{J\Dot{\alpha}}\}=\delta^I_J P_{\alpha\Dot{\alpha}}.
\end{align}
The $R-$symmetry group is $U(\mathcal{N})$ of which $Q_{\alpha}^{I}$ is in the fundamental representation and $\tilde{Q}_{I\Dot{\alpha}}$ is in the anti-fundamental representation. The $R-$symmetry generator carries an adjoint index or equivalently a fundamental and anti-fundamental index. We have,
\begin{align}
    &[R^{I}_{J},Q^K_{\alpha}]=\delta^{K}_{J}Q_{\alpha}^{I}-\frac{1}{\mathcal{N}}\delta^{I}_{J}Q_{\alpha}^{I},\notag\\
    &[R^{I}_{J},\tilde{Q}_{K\Dot{\alpha}}]=\delta^{I}_{K}\tilde{Q}_{J\Dot{\alpha}}-\frac{1}{\mathcal{N}}\delta^{I}_{J}\tilde{Q}_{K\Dot{\alpha}}.
\end{align}
The supersymmetry generators are charges associated to the supersymmetry currents $\mathcal{J}^{I}_{\alpha\Dot{\alpha}\beta}$ and $\mathcal{\tilde{J}_{I\alpha\Dot{\alpha}\Dot{\beta}}}$.
We define the fermionic light ray operators as,
\begin{align}\label{4dfermioniclightrayNeq1}
    &q^I(z,\Bar{z})=\int_{-\infty}^{\infty}dx^{-}~\mathcal{J}_{-\downarrow}(x^-,x^+=0,z,\Bar{z}),\notag\\
    &\tilde{q}_I(z,\Bar{z})=\int_{-\infty}^{\infty}dx^{-}~\mathcal{\tilde{J}}_{-\Dot{\downarrow}}(x^-,x^+=0,z,\Bar{z}),
\end{align}
and the light-ray operator associated to the $R-$symmetry $U(\mathcal{N})$ current,
\begin{align}\label{4dRsymmetrylightrayNeq1}
    r^{I}_{J}(z,\Bar{z})=\int_{-\infty}^{\infty}dx^- J^{I}_{J-}(x^-,x^+=0,z,\Bar{z}).
\end{align}
It is easy to see using  \eqref{4dfermioniclightrayNeq1} and \eqref{4dRsymmetrylightrayNeq1} that both $q^I$ and $\tilde{q}_I$ have twist $2$, as does $r^I_J$.
We now proceed to the computation of the various (anti-)commutators. We also define the shorthand,
\begin{align}
    \delta^2(z_{12})=\delta(\Re{z_1-z_2})\delta(\Im{z_1-z_2})=\delta^2((x^1)_1-(x^1)_2)\delta^2((x^2)_1-(x^2)_2),
\end{align}
when working in Minkowski spacetime $\mathbb{R}^{3,1}$.
\subsubsection*{$\{q,\tilde{q}\}$}
By microcausality we have,
\begin{align}\label{qqtildeeq14d}
    \{q^{I}(z_1,\Bar{z}_1),\tilde{q}_J(z_2,\Bar{z}_2)\}=\delta^2(z_{12})(A_{0,0})^{I}_{J}(z_2,\Bar{z}_2)+\sum_{n=1}^{\infty}\sum_{m=1}^{\infty}\partial_{z_1}^n\partial_{\Bar{z}_1}^m\big(\delta^2(z_{12})\big)(A_{n,m})^I_J(z_2,\Bar{z}_2).
\end{align}
The total twist on the left hand side is $2+2=4$ and thus the $A_{n,m}$ should have twist $2-n-m$. Our assumption is that each $(A_{n,m})^I_J$ is the integral of a local operator of the form $\int_{-\infty}^{\infty}dx^- (\mathcal{O}_{n,m})^I_J(x^-,x^+=0,z,\Bar{z})$. Therefore $(\mathcal{O}_{n,m})^I_J$ has twist $2-n-m$. However, unitarity forbids spinning operators with twist $\tau<2$ in $d=4$. Thus, only $(A_{0,0})^I_J$ is non-zero. To determine its value, integrate \eqref{qqtildeeq14d} over the transverse coordinates $(z_1,\Bar{z}_1)$ which yields,
\begin{align}
    \{Q^I,\tilde{q}_J(z,\Bar{z})\}=(A_{0,0})^I_J(z,\Bar{z})=2\delta^I_J\mathcal{E}(z,\Bar{z}).
\end{align}
We determined this relation using the supersymmetry relation that turns the (anti-holomorphic) supersymmetry current into the stress tensor. Performing another integration then results in,
\begin{align}
    \{Q^I,\tilde{Q}_J\}=2\delta^I_J P_-,
\end{align}
which is the correct relation. Thus we have found,
\begin{align}\label{qqanec4d}
    \{q^I(z_1,\Bar{z}_1),\tilde{q}_J(z_2,\Bar{z}_2)\}=2\delta^2(z_{12})\mathcal{E}(z_2,\Bar{z}_2).
\end{align}
In section \ref{sec:ANEC}, we will study the implications of this equation for the ANEC.
\subsubsection*{$[\mathcal{E},q]$ and $[\mathcal{E},\tilde{q}]$}
By microcausality and unitarity we have,
\begin{align}
    [\mathcal{E}(z_1,\Bar{z}_1),q^I(z_2,\Bar{z}_2)]=\delta^2(z_{12})A(z_2,\Bar{z_2}).
\end{align}
Integrating over $z_1,\Bar{z}_1$ results in,
\begin{align}
    [P_{-},q^I(z_2,\Bar{z}_2]=A(z_2,\Bar{z}_2))=0,
\end{align}
since $q^I(z,\Bar{z})$ is invariant under translations in the $x^-$ direction. Similar arguments show that,
\begin{align}
    [\mathcal{E}(z_1,\Bar{z}_1),\tilde{q}_I(z_2,\Bar{z}_2]=0.
\end{align}

\subsubsection*{$[\mathcal{K},q]$ and $[\mathcal{K},\tilde{q}]$}
Following arguments very similar to the three dimensional case we find,
\begin{align}
    &[\mathcal{K}(z_1,\Bar{z}_1),q^I(z_2,\Bar{z}_2)]=-\frac{i}{2}\delta^{2}(z_{12})q^I(z_2,\Bar{z}_2),\notag\\
    &[\mathcal{K}(z_1,\Bar{z}_1),\tilde{q}_I(z_2,\Bar{z}_2)]=-\frac{i}{2}\delta^2(z_{12})\tilde{q}_{I}(z_2,\Bar{z}_2).
\end{align}

\subsubsection*{$[\mathcal{N}_i,q]$ and $[\mathcal{N}_i,\tilde{q}]$}
These commutators are more involved than the above ones since $\mathcal{N}$ having $\tau=3$ allows for terms involving derivatives of a transverse delta function. Since the transverse space is two dimensional, we can can consider apart from $\partial_i$, the $\epsilon_i^j\partial_j$ structure, where $\epsilon_i^j=\epsilon_{ik}\delta^{kj}$ is the two dimensional Levi-Civita symbol. This is in fact necessary to ensure that $q$ and $\tilde{q}$ have the correct values of the transverse spin. After constraining the commutator, we find,
\begin{align}
    &[\mathcal{N}_i(z_1,\Bar{z}_1),q^{I}(z_2,\Bar{z}_2)]=-i\delta^2(z_{12})\partial_{2i}q^I(z_2,\Bar{z}_2)+i\partial_{1i}\delta^2(z_{12})q^I(z_2,\Bar{z}_2)+\frac{1}{4}\epsilon_i^j\partial_{1i}\delta^2(z_{12})q^I(z_2,\Bar{z}_2),\notag\\
    &[\mathcal{N}_i(z_1,\Bar{z}_1),\tilde{q}_{I}(z_2,\Bar{z}_2)]=-i\delta^2(z_{12})\partial_{2i}\tilde{q}_I(z_2,\Bar{z}_2)+i\partial_{1i}\delta^2(z_{12})\tilde{q}_I(z_2,\Bar{z}_2)-\frac{1}{4}\epsilon_i^j\partial_{1i}\delta^2(z_{12})\tilde{q}_J(z_2,\Bar{z}_2).
\end{align}
As a consistency check, let us verify the Jacobi identity,
\begin{align}\label{Jacobi4dNqqtilde}
    [\mathcal{N}_i(z_1,\Bar{z}_1),\{q^I(z_2,\Bar{z}_2),\tilde{q}_{J}(z_3,\Bar{z}_3)\}]=\{q^I(z_2,\Bar{z}_2),[\mathcal{N}_i(z_1,\Bar{z}_1),\tilde{q}_J(z_3,\Bar{z}_3)]\}+\{\tilde{q}_J(z_3,\Bar{z}_3),[\mathcal{N}_i(z_1,\Bar{z}_1),q^I(z_2,\Bar{z}_2]\}.
\end{align}
The left hand side using the local supersymmetry relation is,
\begin{align}\label{JacobiNwithqqtildeLHS}
    2\delta^I_J\delta^2(z_{23})[\mathcal{N}_i(z_1,\Bar{z}_1),\mathcal{E}(z_3,\Bar{z}_3)]=2\delta^I_J\delta^2(z_{23})\bigg(-i\delta^2(z_{13})\partial_{3i}\mathcal{E}(z_3,\Bar{z}_3)+i\partial_{1i}\delta^2(z_{13})\mathcal{E}(z_3,\Bar{z}_3\big)\bigg).
\end{align}
The right hand side of \eqref{Jacobi4dNqqtilde} contains seven terms,
\begin{align}\label{Jacobi4dNqqtildeRHS}
    &2i\delta^I_J\bigg(-\delta^2(z_{12})\partial_{2i}\delta^2(z_{23})\mathcal{E}(z_3,\Bar{z}_3)+\partial_{1i}\delta^2(z_{12})\delta^2(z_{23})\mathcal{E}(z_3,\Bar{z}_3)-\delta^2(z_{13})\partial_{3i}\delta^2(z_{23})\mathcal{E}(z_3,\Bar{z}_3)\notag\\
    &-\delta^2(z_{13})\delta^2(z_{23})\partial_{3i}\mathcal{E}(z_3,\Bar{z}_3)+\partial_{1i}\delta^2(z_{13})\delta^2(z_{23})\mathcal{E}(z_3,\Bar{z}_3)\notag\\
    &-\frac{1}{4}\epsilon_i^j\partial_{1i}\delta^2(z_{13})\delta^2(z_{23})\mathcal{E}(z_3,\Bar{z}_3)+\frac{1}{4}\epsilon_i^j\partial_{1i}\delta^2(z_{12})\delta^2(z_{23})\mathcal{E}(z_3,\Bar{z}_3)\bigg),
\end{align}
where we note that the $\epsilon_i^j$ terms nicely cancel amongst themselves.
We now use the obvious relation,
\begin{align}
    \delta^2(z_{12})\delta^2(z_{23})=\delta^2(z_{13})\delta^2(z_{23}).
\end{align}
Differentiating both sides of this equation with respect to the position of the second operator results in,
\begin{align}
    &\partial_{2i}\delta^2(z_{12})~\delta^2(z_{23})+\delta^2(z_{12})\partial_{2i}\delta^2(z_{23})=\delta^2(z_{13})\partial_{2i}\delta^2(z_{23})\notag\\&\implies -\partial_{1i}\delta^2(z_{12})~\delta^2(z_{23})+\delta^2(z_{12})\partial_{2i}\delta^2(z_{23})+\delta^2(z_{13})\partial_{3i}\delta^2(z_{23})=0.
\end{align}
Using this identity in \eqref{Jacobi4dNqqtildeRHS}, we find that the first three terms cancel resulting in,
\begin{align}
    2i\delta^2(z_{23})\delta^I_J(-\delta^2(z_{13})\partial_{3i}\mathcal{E}(z_3,\Bar{z}_3)+\partial_{1i}\delta^2(z_{13})\mathcal{E}(z_3,\Bar{z}_3)\big),
\end{align}
which perfectly matches the left hand side \eqref{JacobiNwithqqtildeLHS}.
\subsubsection*{Commutators involving the $r-$charge density}
It is easy to check that the only non-zero commutators involving the $R-$symmetry light ray operator \eqref{4dRsymmetrylightrayNeq1} are with $\mathcal{N}_i$, $q^I$ and $\tilde{q}_I$ (and the obvious one with itself). We find,
\begin{align}
    &[\mathcal{N}_i(z_1,\Bar{z}_1),r^I_J(z_2,\Bar{z}_2)]=-i\delta^2(z_{12})\partial_{2i}r^I_J(z_2,\Bar{z}_2)+i\partial_{1i}\delta^2(z_{12})~r^I_J(z_2,\Bar{z}_2),\notag\\
    &[r^I_J(z_1,\Bar{z}_1),q^K(z_2,\Bar{z}_2)]=\delta^2(z_{12})\big(\delta^K_J q^I(z_2,\Bar{z}_2)-\frac{1}{\mathcal{N}}\delta^I_J q^K(z_2,\Bar{z}_2)\big),\notag\\
    &[r^I_J(z_1,\Bar{z}_1),\tilde{q}_K(z_2,\Bar{z}_2)]=\delta^2(z_{12})\big(\delta^I_K \tilde{q}_J(z_2,\Bar{z}_2)-\frac{1}{\mathcal{N}}\delta^I_J \tilde{q}_K(z_2,\Bar{z}_2)\big).
\end{align}

\subsubsection*{Summary of (anti-)commutation relations}
To summarize, we have found the following relations,
\begin{align}
    &\{q^I(z_1,\Bar{z}_1),\tilde{q}_J(z_2,\Bar{z}_2)\}=2\delta^2(z_{12})\mathcal{E}(z_2,\Bar{z}_2),\notag\\
    &  [\mathcal{E}(z_1,\Bar{z}_1),q^I(z_2,\Bar{z}_2)]=[\mathcal{E}(z_1,\Bar{z}_1),\tilde{q}_I(z_2,\Bar{z}_2]=0,\notag\\
    &[\mathcal{K}(z_1,\Bar{z}_1),q^I(z_2,\Bar{z}_2)]=-\frac{i}{2}\delta^{2}(z_{12})q^I(z_2,\Bar{z}_2),\notag\\
    &[\mathcal{K}(z_1,\Bar{z}_1),\tilde{q}_I(z_2,\Bar{z}_2)]=-\frac{i}{2}\delta^2(z_{12})\tilde{q}_{I}(z_2,\Bar{z}_2),\notag\\
        &[\mathcal{N}_i(z_1,\Bar{z}_1),q^{I}(z_2,\Bar{z}_2)]=-i\delta^2(z_{12})\partial_{2i}q^I(z_2,\Bar{z}_2)+i\partial_{1i}\delta^2(z_{12})q^I(z_2,\Bar{z}_2)+\frac{1}{4}\epsilon_i^j\partial_{1i}\delta^2(z_{12})q^I(z_2,\Bar{z}_2),\notag\\
    &[\mathcal{N}_i(z_1,\Bar{z}_1),\tilde{q}_{I}(z_2,\Bar{z}_2)]=-i\delta^2(z_{12})\partial_{2i}\tilde{q}_I(z_2,\Bar{z}_2)+i\partial_{1i}\delta^2(z_{12})\tilde{q}_I(z_2,\Bar{z}_2)-\frac{1}{4}\epsilon_i^j\partial_{1i}\delta^2(z_{12})\tilde{q}_I(z_2,\Bar{z}_2),\notag\\
    &[\mathcal{N}_i(z_1,\Bar{z}_1),r^I_J(z_2,\Bar{z}_2)]=-i\delta^2(z_{12})\partial_{2i}r^I_J(z_2,\Bar{z}_2)+i\partial_{1i}\delta^2(z_{12})~r^I_J(z_2,\Bar{z}_2),\notag\\
    &[r^I_J(z_1,\Bar{z}_1),q^K(z_2,\Bar{z}_2)]=\delta^2(z_{12})\big(\delta^K_J q^I(z_2,\Bar{z}_2)-\frac{1}{\mathcal{N}}\delta^I_J q^K(z_2,\Bar{z}_2)\big),\notag\\
    &[r^I_J(z_1,\Bar{z}_1),\tilde{q}_K(z_2,\Bar{z}_2)]=\delta^2(z_{12})\big(\delta^I_K \tilde{q}_J(z_2,\Bar{z}_2)-\frac{1}{\mathcal{N}}\delta^I_J \tilde{q}_K(z_2,\Bar{z}_2)\big),\notag\\
    &[r^I_J(z_1,\Bar{z}_1),r^K_L(z_2,\Bar{z}_2]=\delta^2(z_{12})\bigg(\delta^I_L r^{K}_J(z_2,\Bar{z}_2)-\delta^K_J r^{I}_{L}(z_2,\Bar{z}_2)\bigg)
\end{align}
which are to be supplemented by the C{\'o}rdova-Shao bosonic algebra \eqref{4dbosonicfunctionbms}. The remaining commutators are zero.
\subsection{The Algebra of smeared supersymmetric Light-Ray operators}
We now define smeared operators,
\begin{align}
    &\mathcal{Q}^I(f)=\int d^2 z~f(z)q^I(z,\Bar{z}),\notag\\
    &\mathcal{\tilde{Q}}_{I}(\Bar{f})=\int d^2 z~\Bar{f}(\Bar{z})\tilde{q}_I(z,\Bar{z}),\notag\\
    &\rho^I_J(g)=\int d^2 z~g(z,\Bar{z})r^I_J(z,\Bar{z}). 
\end{align}
Note in particular that we have smeared $q^I$ in a holomorphic way and $\tilde{q}_I$ in a anti-holomorphic manner. We have a few (anti-)commutators ahead of us to compute to obtain their algebra. We begin with,
\begin{align}
    \{\mathcal{Q}^I(f),\mathcal{\tilde{Q}}_J(\Bar{f})\}&=\int d^2 z_1~d^2 z_2~f(z_1)\Bar{f}(\Bar{z}_2)\{q^I(z_1,\Bar{z}_1),\tilde{q}_J(z_2,\Bar{z}_2)\}\notag\\
    &=2\delta^I_J\int d^2 z f(z)\Bar{f}(\Bar{z})\mathcal{E}(z,\Bar{z})=2\delta^I_J\mathcal{T}(g),
\end{align}
where $g(z,\Bar{z})=f(z)\Bar{f}(\Bar{z})$. Obvious commutators are,
\begin{align}
    [\mathcal{T}(f),\mathcal{Q}^I(g)]=[\mathcal{T},\mathcal{\tilde{Q}}_I(g)]=0.
\end{align}
Let us next see how the smeared fermionic light-ray operators transform under super-rotations.
\begin{align}
    &[\mathcal{R}_i(Y^i),\mathcal{Q}^I(f)]=\int d^2 z_1\int d^2 z_2\bigg(Y^i(z_1,\Bar{z}_1)f(z_2)[\mathcal{N}_i(z_1,\Bar{z}_1),q^I(z_2,\Bar{z}_2)]\notag\\
    &\qquad\qquad\qquad\qquad\qquad\qquad+\frac{1}{2}\partial_{1i}Y^i(z_1,\Bar{z}_1)f(z_2)[\mathcal{K}(z_1,\Bar{z}_1),q^I(z_2,\Bar{z}_2]\bigg)\notag\\
    &=\int d^2 z_1\int d^2 z_2\bigg(Y^i(z_1,\Bar{z}_1)f(z_2)\big(-i\delta^2(z_{12})\partial_{2i}q^I(z_2,\Bar{z}_2)+i\partial_{1i}\delta^2(z_{12})q^I(z_2,\Bar{z}_2)\notag\\
    &+\frac{\epsilon_i^j}{4}\partial_{1j}\delta^2(z_{12})q^I(z_2,\Bar{z}_2)\big)-\frac{i}{4}\partial_{1i}Y^i(z_1,\Bar{z}_1)f(z_2)\delta^2(z_{12})q^I(z_2,\Bar{z}_2)\bigg)\notag\\
    &=i\int d^2 z~g(z,\Bar{z})q^I(z,\Bar{z})=i\mathcal{Q}^I(g),
\end{align}
where $g(z,\Bar{z})=Y^i(z,\Bar{z})\partial_i f(z)-\frac{1}{4}(\partial_i Y^i(z,\Bar{z}))f(z)+\frac{i\epsilon_i^j}{4}(\partial_j Y^j(z,\Bar{z}))f(z)$. It might seem like $g$ is not a holomorphic function of $z$ given this formula, but we shall see that it is actually a holomorphic function in disguise if we choose $Y^i$ to be either holomorphic or anti-holomorphic which ensures that the algebra closes nicely, yielding the usual supersymmetric $\mathfrak{bms}_4$ algebra.

Next, consider the transformation of $\mathcal{\tilde{Q}}_I$ under super-rotations.
\begin{align}
    &[\mathcal{R}_i(Y^i),\mathcal{\tilde{Q}}_I(\Bar{f})]=\int d^2 z_1\int d^2 z_2\bigg(Y^i(z_1,\Bar{z}_1)\Bar{f}(\Bar{z}_2)[\mathcal{N}_i(z_1,\Bar{z}_1),q^I(z_2,\Bar{z}_2)]\notag\\
    &\qquad\qquad\qquad\qquad\qquad\qquad+\frac{1}{2}\partial_{1i}Y^i(z_1,\Bar{z}_1)\Bar{f}(\Bar{z}_2)[\mathcal{K}(z_1,\Bar{z}_1),\tilde{q}_I(z_2,\Bar{z}_2]\bigg)\notag\\
    &=\int d^2 z_1\int d^2 z_2\bigg(Y^i(z_1,\Bar{z}_1)\Bar{f}(\Bar{z}_2)\big(-i\delta^2(z_{12})\partial_{2i}\tilde{q}_I(z_2,\Bar{z}_2)+i\partial_{1i}\delta^2(z_{12})q^I(z_2,\Bar{z}_2)\notag\\
    &-\frac{\epsilon_i^j}{4}\partial_{1j}\delta^2(z_{12})\tilde{q}_I(z_2,\Bar{z}_2)\big)-\frac{i}{4}\partial_{1i}Y^i(z_1,\Bar{z}_1)\Bar{f}(\Bar{z}_2)\delta^2(z_{12})\tilde{q}_I(z_2,\Bar{z}_2)\bigg)\notag\\
    &=i\int d^2 z~\Bar{g}(z,\Bar{z})\tilde{q}_I(z,\Bar{z})=i\mathcal{\tilde{Q}}_I(\Bar{g}),
\end{align}
where $\Bar{g}(z,\Bar{z})=Y^i(z,\Bar{z})\partial_i \Bar{f}(\Bar{z})-\frac{1}{4}(\partial_i Y^i(z,\Bar{z}))\Bar{f}(\Bar{z})-\frac{i\epsilon_i^j}{4}(\partial_j Y^j(z,\Bar{z}))\Bar{f}(\Bar{z})$. At first glance, $\Bar{g}$ does not look like an anti-holomorphic function but as we shall see, it actually is one once we specify the holomorphicity of the super-rotations.
Finally, we should compute the non-trivial commutators involving the smeared $R-$symmetry light-ray operator $\rho^I_J(g)$. We find for instance,
\begin{align}
    &[\rho^I_J(g(z,\Bar{z})),\mathcal{Q}^K(f(z))]=\delta^K_J\mathcal{Q}^I(h)-\frac{\delta^I_J}{\mathcal{N}}\mathcal{Q}^K(h),\notag\\
    &[\rho^I_J(g(z,\Bar{z})),\mathcal{\tilde{Q}}_K(f(z))]=\delta^I_K\mathcal{\tilde{Q}}_J(h)-\frac{\delta^I_J}{\mathcal{N}}\mathcal{\tilde{Q}}_K(\Bar{h}),
\end{align}
where $h=g(z,\Bar{z})f(z)$ and $\Bar{h}=g(z,\Bar{z})\Bar{f}(\Bar{z})$.

However, this immediately results in an inconsistency since the smearing function for the $R-$symmetry light-ray operator is a function of $z$ and $\Bar{z}$ whereas $\mathcal{Q}^I$ and $\mathcal{\tilde{Q}}_I$ are smeared in holomorphic and anti-holomorphic ways respectively. There is absolutely no way for these functions to be holomorphic and anti-holomorphic for general $g(z,\Bar{z})$. The only possibility is that $g(z,\Bar{z})$ is a constant and that corresponds to just the global $R-$symmetry generator $R^I_J$. Therefore, the general smeared $R-$symmetry light-ray operator is not part of this algebra. A similar observation was made in \cite{Banerjee:2022lnz} where the authors show that the $SU(8)$ global $R-$symmetry of $\mathcal{N}=8$ supergravity in four dimensional flat space, admits no infinite dimensional extension at null infinity. What we have shown here is that for any $\mathcal{N}-$extended superconformal field theories, the symmetry algebra of our light-ray operators at a null hypersurface does not include an infinite dimensional extension of the $R-$symmetry. Since we can map our null hypersurface at $x^+=0$ to $\mathcal{I}^+$ thanks to conformal symmetry, we provide another (generalized) proof of their statement. This is in contrast to $d=3$, where we found a consistent infinite dimensional algebra that included the local $R-$symmetry generators in section \ref{sec:3d}. 

However, the \textit{generalized} supersymmetric $\mathfrak{bms}_4$ algebra does indeed close even after including the local $R-$symmetry generators. This requires us to smear $\mathcal{Q}$ and $\mathcal{\tilde{Q}}$ with generic functions $f(z,\Bar{z})$ (and to be democratic, the same for super-rotations). This results in a generalized $\mathcal{N}-$ extended supersymmetric $\mathfrak{bms}_4$ algebra that includes the local $R-$symmetry tower, resulting in a much larger algebra. That being said, we now proceed with the holomorphically smeared supersymmetric light-ray operators to draw easy comparison to the more familiar supersymmetric $\mathfrak{bms}_4$ algebras. The important point that we wish to emphasize here is that our SCFT construction allows for the generalized version of the algebra.

Thus, in addition to \eqref{4dbosonicfunctionbms}, we find the following (anti-)commutators that form a supersymmetric $\mathfrak{bms}_4$ algebra. 
\begin{align}\label{4dsusyfunctionbms}
    &[\mathcal{T}(f_1),\mathcal{T}(f_2)]=0,~~[\mathcal{T}(f_1),\mathcal{R}(Y)]=i\mathcal{T}(g),~[\mathcal{R}(Y_1),\mathcal{R}(Y_2)]=i\mathcal{R}(Y_3),\notag\\
    &\{\mathcal{Q}^I(\epsilon_1),\mathcal{\tilde{Q}}_J(\epsilon_2)\}=2\delta^I_J\mathcal{T}(\epsilon_1\epsilon_2),~[\mathcal{R}(Y),\mathcal{Q}^I(f)]=i\mathcal{Q}^I(h),~[\mathcal{R}(Y),\mathcal{\tilde{Q}}_I(\Bar{f})]=i\mathcal{Q}^I(\Bar{h}),
\end{align}
where,
\begin{align}\label{4dsusyfunctionbmscont}
    &g(z,\Bar{z})=\frac{1}{2}(\partial_i Y^i(z,\Bar{z}))f_1(z,\Bar{z})-Y^i\partial_i f_1(z,\Bar{z}),\notag\\&Y_3^i(z,\Bar{z})=Y_1^j(z,\Bar{z})\partial_j Y_2^i(z,\Bar{z})-(\partial_j Y_1^i(z,\Bar{z}))Y_2^j(z,\Bar{z}),\notag\\ &h(z,\Bar{z})=Y^i(z,\Bar{z})\partial_i f(z)-\frac{1}{4}(\partial_i Y^i(z,\Bar{z}))f(z)+\frac{i\epsilon_i^j}{4}(\partial_j Y^j(z,\Bar{z}))f(z),\notag\\
    &\Bar{h}(z,\Bar{z})=Y^i(z,\Bar{z})\partial_i \Bar{f}(\Bar{z})-\frac{1}{4}(\partial_i Y^i(z,\Bar{z}))\Bar{f}(\Bar{z})-\frac{i\epsilon_i^j}{4}(\partial_j Y^j(z,\Bar{z}))\Bar{f}(\Bar{z})
\end{align}

\subsection{The $\mathcal{N}$-extended supersymmetric $\mathfrak{bms}_4$ mode algebra}
We now introduce modes analogous to \eqref{4dbosonicmodes}. 
\begin{align}
    F^I_r=e^{\frac{i\pi}{2}}\mathcal{Q}^I(z^r),\tilde{F}_{I,r}=e^{\frac{i\pi}{2}}\tilde{Q}_I(\Bar{z}^r).
\end{align}
where $r\in \mathbb{Z}+\frac{1}{2}$ or $r\in\mathbb{Z}$ and $m,n\in\mathbb{Z}$.

Using this along with  \eqref{4dsusyfunctionbms} and \eqref{4dsusyfunctionbmscont}, we can then compute the algebra of the modes. Let us start with $\{F_r^I,\tilde{F}_{J,s}\}$. We have,
\begin{align}
    \{F_r^I,\tilde{F}_{J,s}\}=e^{\frac{i\pi}{2}}\{\mathcal{Q}^I(z^r),\mathcal{\tilde{Q}}_J(\Bar{z}^s)\}=2i\delta^I_J\mathcal{T}(z^r\Bar{z}^s)=2\delta^I_JM_{r,s}.
\end{align}
Next, we compute,
\begin{align}
    [L_n,F_r^I]=e^{\frac{3i\pi}{4}}[\mathcal{R}(Y^i=-\delta^i_z z^{n+1}),\mathcal{Q}^I(z^r)]=e^{\frac{3i\pi}{4}}\mathcal{Q}(h),
\end{align}
where using \eqref{4dsusyfunctionbmscont} we find,
\begin{align}
    h(z,\Bar{z})&=-\delta^i_z z^{n+1}\partial_i z^r-\frac{1}{4}\partial_i(-\delta^i_z z^{n+1})z^r+i\frac{\epsilon_i^j}{4}\partial_j(-\delta^j_z z^{n+1})z^r\notag\\
    &=-r z^{n+r}+\frac{n+1}{4}z^{n+r}+\frac{n+1}{4}z^{n+r}=\bigg(\frac{n+1}{2}-r\bigg)z^{n+r}.
\end{align}
Thus,
\begin{align}
    [L_n,F_r^I]=\bigg(\frac{n+1}{2}-r\bigg)F_{r+n}^{I}.
\end{align}
Next we compute the potentially problematic commutator,
\begin{align}
    [\Bar{L}_n,F_r^I]=e^{\frac{3i\pi}{4}}[\mathcal{R}(Y^i=-i\delta^i_{\Bar{z}}\Bar{z}^{n+1}),\mathcal{Q}^I(z^r)]=e^{\frac{3i\pi}{4}}\mathcal{Q}^I(h),
\end{align}
where now,
\begin{align}
    h(z,\Bar{z})&=-\delta^i_{\Bar{z}}\Bar{z}^{n+1}\partial_i z^r-\frac{1}{4}\partial_i(-\delta^i_{\Bar{z}}\Bar{z}^{n+1})z^r+\frac{i\epsilon_i^j}{4}\partial_j(-\delta^j_{\Bar{z}}\Bar{z}^{n+1})z^r\notag\\
    &=\frac{(n+1)}{4}\Bar{z}^{n}z^r+\frac{(-i)^2}{4}\Bar{z}^nz^r=0.
\end{align}
Therefore,
\begin{align}
    [\Bar{L}_n,F_r^I]=0.
\end{align}
As we discussed earlier, we see that the terms in $h$ rearrange themselves to ensure that $F_r^I$ are holomorphic modes and do not mix with the anti-holomorphic sector under super-rotations generated by $\Bar{L}_n$ which are purely anti-holomorphic. Similarly, one can go ahead and compute the commutators involving $\tilde{F}_{I,r}$. The final result is the $\mathcal{N}-$extended supersymmetric $\mathfrak{bms}_4$ algebra \cite{Fotopoulos:2020bqj}.

\begin{4dBMSNsusy}
\begin{align}\label{4dbosonicbms}
    &\qquad\qquad\qquad\qquad\qquad\qquad[M_{m,n},M_{m',n'}]=0,\notag\\
    &[L_n,M_{m,p}]=\bigg(m-\frac{(n+1)}{2}\bigg)M_{r+m,p},~~[\Bar{L}_n,M_{m,p}]=\bigg(p-\frac{(n+1)}{2}\bigg)M_{m,p+n},\notag\\
    &[L_m,L_n]=(m-n)L_{m+n},\qquad[L_m,\Bar{L}_m]=0,\qquad[\Bar{L}_m,\Bar{L}_n]=(m-n)\Bar{L}_{m+n},\notag\\&[F^I_r,M_{m,n}]=0,\qquad\qquad\qquad\{F^I_r,\tilde{F}_{J,s}\}=2\delta^I_JM_{r+s},\qquad[\tilde{F}_{I,r},M_{m,n}]=0,\notag\\
    &[L_n,F_r^I]=\big(r-\frac{(n+1)}{2}\big)F_{r+n}^I,\qquad\qquad\qquad[\Bar{L}_n,\tilde{F}_{I,r}]=\big(r-\frac{n+1}{2}\big)\tilde{F}_{I,r+n},\notag\\
    &[\Bar{L}_n,F_r^I]=0,\qquad\qquad\qquad\qquad\qquad\qquad\qquad[L_n,\tilde{F}_{I,r}]=0.
\end{align}
\end{4dBMSNsusy}
Therefore, any $\mathcal{N}-$extended super-conformal field theory in four dimensions contains supersymmetric light-ray operators which contain a sub-algebra that is the $\mathcal{N}-$extended supersymmetric $\mathfrak{bms}_4$ algebra.

\section{A Supersymmetric proof of ANEC}\label{sec:ANEC}
The essential point in our construction is the existence of a local square root of the averaged null energy operator \eqref{qqANEC3d}, \eqref{qqanec4d}. The averaged null energy condition is the statement that this operator has a positive semi-definite expectation value in any state of a unitary quantum field theory. In two dimensions, the ANEC follows quite simply from the spectral condition since it can be identified with the null component of the momentum generator \cite{Hartman:2023ccw}. If we consider supersymmetric theories, then it becomes all the more simpler since the null component of the momentum generator is the anti-commutator of supercharges, which trivializes its positivity. This is in contrast with the story in higher dimensions, where as we discussed in the introduction, it has been  proven using information theoretic ideas \cite{Faulkner:2016mzt} and conformal bootstrap methods \cite{Hartman:2016lgu}. Given the supersymmetric light-ray construction we have been discussing, it is natural to see if the ANEC follows from its existence, similar to the argument in two dimensional supersymmetric theories, just with a local algebra replacing a global one. We show this explicitly in this section, deriving a simple, elementary proof of the ANEC in supersymmetric conformal field theories in $d=3,4$.
\subsection{Three dimensions}
We have in $d=3$ with $\mathcal{N}=1$ supersymmetry,
\begin{align}
    \{q(y_1),q(y_2)\}=2\delta(y_1-y_2)\mathcal{E}(y_2),
\end{align}
with the smeared versions satisfying,
\begin{align}
    \{\mathcal{Q}(f_1),\mathcal{Q}(f_2)\}=2\mathcal{T}(f_1f_2).
\end{align}
If we take $f_1=f_2=f$ we obtain,
\begin{align}\label{Qf2vsTf2}
     &\{\mathcal{Q}(f),\mathcal{Q}(f)\}=2\mathcal{T}(f^2)\implies \mathcal{Q}(f)^2=\mathcal{T}(f^2)=\int_{-\infty}^{\infty}dy~ f(y)^2 \mathcal{E}(y).
\end{align}
Now consider some state in the SCFT Hilbert space $|\Psi\rangle$ and form,
\begin{align}\label{operatornormpositive}
    \langle \Psi|\mathcal{Q}(f)^2|\Psi\rangle=||\mathcal{Q}(f)|\Psi\rangle||^2\ge 0,
\end{align}
since this is the square of the norm of the state. Using \eqref{Qf2vsTf2}, we can re-write this as,
\begin{align}
    \langle \Psi|\mathcal{Q}(f)^2|\Psi\rangle=\int_{-\infty}^{\infty}dy~f(y)^2\langle \Psi|\mathcal{E}(y)|\Psi\rangle\ge 0,
\end{align}
where we used \eqref{operatornormpositive}. Since $f(y)$ is an arbitrary function which we can take to be sufficiently localized, this inequality must hold for the expectation value inside the integral thus resulting in,
\begin{align}
    \langle \Psi|\mathcal{E}(y)|\Psi\rangle\ge 0,
\end{align}
which is the averaged null energy condition. Therefore, supersymmetry, microcausality and unitarity imply the ANEC in a SCFT. A similar proof can be extended for theories with higher supersymmetry. 
\subsection{Four dimensions}
It is also simple to extend the construction to higher dimensions. For instance, consider $\mathcal{N}=1$ supersymmetry in $d=4$. We have,
\begin{align}
    \{q^I(z_1,\Bar{z}_1),\tilde{q}_J(z_2,\Bar{z}_2)\}=2\delta^I_J\delta^2(z_{12})\mathcal{E}(z_2,\Bar{z}_2).
\end{align}
The smeared version is,
\begin{align}
    \{\mathcal{Q}^I(f_1),\mathcal{\tilde{Q}}_J(f_2)\}=2\delta^I_J\mathcal{T}(f_1 f_2).
\end{align}
We now take $f_1=f(z)$ and $f_2=f(z)^*$.
\begin{align}
    \{\mathcal{Q}^I(f),\mathcal{\tilde{Q}}_J(f(z)^*)\}=2\delta^I_J\mathcal{T}(f^2).
\end{align}
We consider a state $|\Psi\rangle\in\mathcal{H}$ and construct,
\begin{align}
    \langle \Psi|\mathcal{Q}^I(f)\mathcal{\tilde{Q}}_J(f^*)+\mathcal{\tilde{Q}}_J(f^*)\mathcal{Q}^I(f)|\Psi\rangle=2\delta^I_J\langle \Psi|\mathcal{T}(|f|^2)|\Psi\rangle.
\end{align}
Using the fact that $\mathcal{\tilde{Q}}_J(f^*)=(\mathcal{Q}^J(f))^\dagger$ and linearity of quantum mechanics, we can re-write the above expectation value as,
\begin{align}
    \langle \Psi|(\mathcal{\tilde{Q}}_I(f^*))^\dagger\mathcal{\tilde{Q}}_J(f^*)|\Psi\rangle+\langle \Psi|(\mathcal{Q}_J(f))^\dagger\mathcal{Q}^I(f)|\Psi\rangle.
\end{align}
Taking $I=J$ to be some fixed value of the index then gives us,
\begin{align}
    &||\mathcal{\tilde{Q}}_I(f)|\Psi\rangle||^2+||\mathcal{Q}^I(f)|\Psi\rangle||^2\ge 0\notag\\
    &\implies \langle \Psi|\mathcal{T}(f^2)|\Psi\rangle=\int d^2 z |f(z)|^2  \langle \Psi|\mathcal{E}(z,\Bar{z})|\Psi\rangle\ge 0.
\end{align}
Taking the function $f(z)$ to be sufficiently localized then results in the averaged null energy condition,
\begin{align}
    \langle \Psi|\mathcal{E}(z,\Bar{z})|\Psi\rangle\ge 0,
\end{align}
which completes the proof.

\acknowledgments
I would like to thank Abhijit Gadde, Shiraz Minwalla, Sridip Pal, Priyadarshini Pandit, Onkar Parrikar, Harshit Rajgadia, Pratik Rath and Amit Suthar for discussions and very useful comments. This work was supported by the Department of Atomic Energy, Government of India, under Project Identification Number RTI-4012 and the Infosys Endowment for the study of the Quantum Structure of Spacetime.

\appendix
\section{Notation and Conventions}\label{app:Notation}
We outline our notation and conventions for indices, vectors, spinors, gamma matrices and light-cone coordinates in this appendix. 
\subsection{Three dimensions}
The Lorentz group in $d=3$ is $SO(2,1)$ which is homomorphic to $SL(2,\mathbb{R})$. Vector indices are denoted by Greek letters $\mu,\nu,\cdots$ whereas fundamental spinor indices $a,b,\cdots$ are represented in Latin.
The flat $\mathbb{R}^{2,1}$ metric is given by,
\begin{align}
    ds^2=\eta_{\mu\nu}dx^{\mu}dx^{\nu}=-dt^2+dx^2+dy^2=-dx^{+}dx^{-}+dy^2,
\end{align}
where our light-cone coordinates are,
\begin{align}
    x^{\pm}=t\pm x.
\end{align}
The non-zero metric components are,
\begin{align}
    \eta_{-+}=\eta_{+-}=-\frac{1}{2},\eta_{yy}=1.
\end{align}
Vector indices are raised and lowered using the metric $\eta_{\mu\nu}$ and its inverse $\eta^{\mu\nu}$.
\begin{align}
    A^\mu=\eta^{\mu\nu}A_\nu, A_\mu=\eta_{\mu\nu}A^\mu.
\end{align}
The non-zero components of the inverse metric are,
\begin{align}
    \eta^{-+}=\eta^{+-}=-2,\eta^{yy}=1.
\end{align}
Given a vector index, we can trade it for a pair of spinor indices using the Gamma matrices,
\begin{align}
    &(\gamma^t)_a^b=\begin{pmatrix}
        0&-1\\1&0
    \end{pmatrix},~(\gamma^x)_a^b=\begin{pmatrix}
        0&1\\1&0
    \end{pmatrix},~(\gamma^y)_a^b=\begin{pmatrix}
        1&0\\0&-1
    \end{pmatrix},\notag\\
    &\qquad\qquad(\gamma^+)_a^b=\begin{pmatrix}
        0&0\\2&0
    \end{pmatrix}, (\gamma^-)_a^b=\begin{pmatrix}
        0&-2\\0&0
    \end{pmatrix},
\end{align}
which satisfy the Clifford algebra,
\begin{align}
    \{\gamma^\mu,\gamma^\nu\}_a^b=2\delta_a^b\eta^{\mu\nu}.
\end{align}
This can easily be verified using the multiplication rule that we obtain by virtue of the Pauli matrix algebra:
\begin{align}
    (\gamma^\mu)^c_a (\gamma^\nu)^b_c=\delta^b_a \eta^{\mu\nu}+\epsilon^{\mu\nu\rho}(\gamma_\rho)^b_a.
\end{align}
The spinor indices can be raised and lowered using the Levi-Civita symbol $\epsilon^{ab}$. We choose,
\begin{align}
    \epsilon_{ab}=\epsilon^{ab}=\begin{pmatrix}
        0&1\\-1&0
    \end{pmatrix}.
\end{align}
Then,
\begin{align}
    x^a=\epsilon^{ab}x_b, x_a=\epsilon_{ba}x^b.
\end{align}
Let us now consider the translation generator $P_{\mu}$. We can trade it for a bi-spinor by contracting with the Pauli matrices (we find it useful to work with both indices below or above since they are symmetric),
\begin{align}\label{PmatrixApp}
    &P_{ab}=P_{\mu}(\gamma^\mu)_{ab}=\begin{pmatrix}
        2P_{-}&P_{y}\\
        P_{y}& 2P_{+}.
    \end{pmatrix},~  P^{ab}=P_{\mu}(\gamma^\mu)^{ab}=\begin{pmatrix}
        2P_{+}&-P_{y}\\
        -P_{y}& 2P_{-}.
    \end{pmatrix},
\end{align}
where the light-cone momenta components are,
\begin{align}
   P_{\pm}=\frac{(P_t\pm P_x)}{2}.
\end{align}
Finally, we denote the spinor components in the ``light-cone" basis as,
\begin{align}
    \lambda_1=\lambda_{\downarrow},\lambda_2=\lambda_{\uparrow},
\end{align}
and the index-raised versions,
\begin{align}
    \lambda^{\downarrow}=\lambda_{\uparrow},\lambda^{\uparrow}=-\lambda_{\downarrow}.
\end{align}
For example, we have for $P_{ab}$ defined in \eqref{PmatrixApp},
\begin{align}
    P_{\downarrow\downarrow}=2P_-,P_{\downarrow\uparrow}=P_{\uparrow\downarrow}=P_y,P_{\uparrow\uparrow}=2P_+.
\end{align}
Thus, colloquially speaking, $\downarrow$ and $\uparrow$ are the square roots of the vector light-cone directions $-$ and $+$ respectively with their ``product" giving the $y$ direction.

\subsection{Four dimensions}
The flat four dimensional Minkowski metric is,
\begin{align}
    ds^2=-(dx^0)^2+(dx^1)^2+(dx^2)^2+(dx^3)^2=-dx^{+}dx^{-}+dz d\Bar{z},
\end{align}
where $x^{\pm}=x^0\pm x^{3}$ and $z=x^{1}+ i x^{2}$, $\Bar{z}=x^{1}-i x^{2}$. $z$ and $\Bar{z}$ are complex conjugates of each other since we are in Lorentzian signature. The Lorentz group  is semi-simple and up to a suitable choice of reality condition for Minkowski space, is homomorphic to $SL(2)_L\times SL(2)_R$. Vector indices are denoted by greek letters $\mu,\nu,\cdots$, left handed spinor indices by $\alpha,\beta,\cdots$ and right handed spinor indices by $\Dot{\alpha},\Dot{\beta},\cdots$. A spacetime vector can be traded for a bi-spinor using,
\begin{align}
    p_{\alpha\Dot{\alpha}}=p_\mu(\sigma^\mu)_{\alpha\Dot{\alpha}}=\begin{pmatrix}
        p_0-p_{3}&-p_{1}+i p_{2}\\
    -p_1-ip_2&p_0+p_{3}\end{pmatrix}=\begin{pmatrix}
        2p_-& -2p_z\\
        -2p_{\Bar{z}}&2p_{+}
    \end{pmatrix},
\end{align}
where our four dimensional sigma matrix convention is,
\begin{align}
    &\sigma^0=\begin{pmatrix}
        1&0\\0&1
    \end{pmatrix},\sigma^1=\begin{pmatrix}
        0&-1\\-1&0
    \end{pmatrix},\sigma^2=\begin{pmatrix}
        0&i\\-i&0
    \end{pmatrix},\sigma^3=\begin{pmatrix}
        -1&0\\0&1
    \end{pmatrix}.
\end{align}
The barred sigma matrices are defined in the usual way,
\begin{align}
    &(\Bar{\sigma}^\mu)^{\Dot{\alpha}\alpha}=\epsilon^{\alpha\beta}\epsilon^{\Dot{\alpha}\Dot{\beta}}(\sigma^\mu)_{\beta\Dot{\beta}},\notag\\
    &\implies \Bar{\sigma}^0=\sigma^0,\Bar{\sigma}^i=-\sigma^i,i=1,2,3.
\end{align}
We have,
\begin{align}
    p^{\Dot{\alpha}\alpha}=\begin{pmatrix}
        2p_+&2p_z\\
        2p_{\Bar{z}}&2p_-
    \end{pmatrix}.
\end{align}
Exactly as in $d=3$, we denote the spinor components as $\lambda_1=\lambda_{\downarrow}$ and $\lambda_2=\lambda_{\uparrow}$ and similarly for the dotted indices. For instance we have for a vector $P_{\downarrow\Dot{\downarrow}}=2P_{-}$.

\section{Proof of (anti-)commutation relations in free theory}\label{app:FreeTheory}
In this appendix, we test our supersymmetric light-ray algebra in the simple setting of free supersymmetric conformal field theories. We work in the framework of light-cone quantization and use the canonical (anti-)commutation relations to check our results. All composite operators we define below are implicitly normal ordered.
\subsection{Three dimensions}
Consider the three dimensional free $\mathcal{N}=1$ Wess-Zumino action,
\begin{align}
    S=\int d^3 x~\bigg(-\frac{1}{2}\eta^{\mu\nu}\partial_\mu\phi\partial_\nu \phi-\frac{i}{2}\psi^a \slashed{\partial}_{ab}\psi^b\bigg).
\end{align}
$\phi$ is a real scalar field and $\psi^a$ is a Majorana fermion. The equations of motion are,
\begin{align}
    &\qquad\qquad\qquad\Box\phi=0,\notag\\
    &2\partial_-\psi_{\uparrow}-\partial_y\psi_{\downarrow}=0,\partial_y \psi_{\uparrow}-2\partial_{+}\psi_{\downarrow}=0.
\end{align}
Thus, there is one propagating scalar degree of freedom and one fermionic degree of freedom (since $\psi_{\uparrow}$ can be solved in terms of $\psi_{\downarrow}$ using the constraint equation). We have also used the equation of motion to set the auxiliary field to zero.

We quantize this theory on the light-front at $x^{+}=0$. The canonical (anti-)commutation relations are,
\begin{align}
    &[\phi(x_1^-,0,y_1),\partial_{2-}\phi(x_2^-,0,y_2)]=\frac{i}{2}\delta(x_1^- -x_2^-)\delta(y_1-y_2),\notag\\
    &\{\psi_{\downarrow}(x_1^-,0,y_1),\psi_{\downarrow}(x_2^-,0,y_2)\}=\delta(x_1^- - x_2^-)\delta(y_1-y_2).
\end{align}
The operators of interest to us are those in the stress tensor multiplet. These are the supersymmetry current and the stress tensor itself. The latter is given by,
\begin{align}
    T_{\mu\nu}&=\bigg(\partial_\mu\phi\partial_\nu\phi-\frac{\eta_{\mu\nu}}{2}(\partial \phi)^2+\frac{1}{8}(\eta_{\mu\nu}\Box-\partial_\mu\partial_\nu)(\phi^2)\bigg)\notag\\
    &-\frac{i}{8}\bigg(\psi^a(\sigma_\mu)_{ab}\partial_\nu \psi^b-\partial_\nu\psi^a(\gamma_\mu)_{ab}\psi^b+\psi^a(\gamma_\nu)_{ab}\partial_\mu \psi^b-\partial_\mu\psi^a(\gamma_\nu)_{ab}\psi^b\bigg).
\end{align}
Its $(--)$ component is thus,
\begin{align}
    T_{--}=(\partial_-\phi)^2-\frac{1}{8}(\partial_-)^2(\phi^2)+\frac{i}{2}\psi_{\downarrow}\partial_-\psi_{\downarrow}.
\end{align}
The averaged null energy operator is then given by,
\begin{align}
    \mathcal{E}(y)=\int_{-\infty}^{\infty}dx^-\bigg((\partial_-\phi)^2+\frac{i}{2}\psi_{\downarrow}\partial_{-}\psi_{\downarrow}\bigg),
\end{align}
where all fields are located at $x^+=0$.
The supersymmetry current takes the form,
\begin{align}
    \mathcal{J}_{abc}=\frac{1}{2\sqrt{2}}\bigg((\psi_a\partial_{bc}\phi+\psi_b\partial_{ca}\phi+\psi_c\partial_{ab}\phi)-\frac{1}{3}\big(\partial_{bc}\psi_a~\phi+\partial_{ca}\psi_b~\phi+\partial_{ab}\psi_c~\phi\big)\bigg).
\end{align}
One can check that this quantity is traceless with respect to any pair of spinor indices and is also conserved on-shell. We find for its $\downarrow\downarrow\downarrow$ component,
\begin{align}
    \mathcal{J}_{\downarrow\downarrow\downarrow}=\frac{3}{2\sqrt{2}}\bigg(\psi_{\downarrow}\partial_{\downarrow\downarrow}\phi-\frac{1}{3}\partial_{\downarrow\downarrow}\psi_{\downarrow}~\phi\bigg).
\end{align}
The supersymmetric $q$ light-ray operator is then given by,
\begin{align}
    q(y)=\sqrt{2}\int_{-\infty}^{\infty}dx^-\bigg(\psi_{\downarrow}\partial_{-}\phi\bigg),
\end{align}
with the fields inserted at $x^{+}=0$. Let us now compute the anti-commutator $\{q(y_1),q(y_2)\}$ and verify that it equals $2\delta(y_1-y_2)\mathcal{E}(y_2)$. We have,
\begin{align}\label{freetheoryqqstep1}
    \{q(y_1),q(y_2)\}=2\int_{-\infty}^{\infty}dx_1^-\int_{-\infty}^{\infty}dx_2^-\{\psi_{\downarrow}(x_1^-,0,y_1)\partial_{1-}\phi(x_1^-,0,y_1),\psi_{\downarrow}(x_2^-,0,y_2)\partial_{2-}\phi(x_2^-,0,y_2)\}.
\end{align}
We now use the following identity. Let $F_1,F_2$ be fermionic operators and $B_1,B_2$ be bosonic. Then,
\begin{align}
    \{F_1 B_1,F_2 B_2\}=F_1 F_2[B_1,B_2]+\{F_2,F_1\}B_2 B_1,
\end{align}
which can be verified by a direct expansion. Thus we have,
\begin{align}
    &\{\psi_{\downarrow}(x_1^-,0,y_1)\partial_{1-}\phi(x_1^-,0,y_1),\psi_{\downarrow}(x_2^-,0,y_2)\partial_{2-}\phi(x_2^-,0,y_2)\}\notag\\&=\psi_{\downarrow}(x_1^-,0,y_1)\psi_{\downarrow}(x_2^-,0,y_2)[\partial_{1-}\phi(x_1^-,0,y_1),\partial_{2-}\phi(x_2^-,0,y_2)]\notag\\
    &+\{\psi_{\downarrow}(x_2^-,0,y_2),\psi_{\downarrow}(x_1^-,0,y_1)\}\partial_{2-}\phi(x_2^-,0,y_2)\partial_{1-}\phi(x_1^-,0,y_1)\notag\\
    &=\frac{i}{2}\psi_{\downarrow}(x_1^-,0,y_1)\psi_{\downarrow}(x_2^-,0,y_2)\partial_{1-}\delta(x_1^- -x_2^-)\delta(y_1-y_2)\notag\\&+\delta(x_2^- -x_1^-)\delta(y_1-y_2)\partial_{1-}\phi(x_1^-,0,y_1),\partial_{2-}\phi(x_2^-,0,y_2).
\end{align}
Substituting this into \eqref{freetheoryqqstep1}, integrating by parts, we find,
\begin{align}
    \{q(y_1),q(y_2)\}&=2\delta(y_1-y_2)\int_{-\infty}^{\infty}dx^-\bigg(-\frac{i}{2}\partial_-\psi_{\downarrow}(x^-,0,y)~\psi_{\downarrow}(x^-,0,y)+\big(\partial_{-}\phi(x^-,0,y)\big)^2\bigg)\notag\\
    &=2\delta(y_1-y_2)\int_{-\infty}^{\infty}\bigg((\partial_-\phi)^2+\frac{i}{2}\psi_{\downarrow}\partial_-\psi_{\downarrow}\bigg)\notag\\
    &=2\delta(y_1-y_2)\mathcal{E}(y_2),
\end{align}
which is exactly the result we derived using symmetry and unitarity for general theories. To conclude this section, let us compute the remaining commutators between $q(y)$ and the operators $\mathcal{K}(y)$ and $\mathcal{N}_y(y)$. In the free theory we have,
\begin{align}
    \mathcal{K}(y)=\int_{-\infty}^{\infty}dx^- x^- T_{--}(x^-,0,y)=\int_{-\infty}^{\infty}dx^-~x^-\bigg((\partial_-\phi)^2+\frac{i}{2}\psi_{\downarrow}\partial_{-}\psi_{\downarrow}\bigg),
\end{align}
We then have,
\small
\begin{align}
    &[\mathcal{K}(y_1),q(y_2)]\notag\\&=\sqrt{2}\int_{-\infty}^{\infty}dx_1^-\int_{-\infty}^{\infty}dx_2^-~x_1^-[(\partial_{1-}\phi(x_1^-,0,y_1))^2+\frac{i}{2}\psi_{\downarrow}(x_1^-,0,y_1)\partial_{1-}\psi_{\downarrow}(x_1^-,0,y_1),\psi_{\downarrow}(x_2^-,0,y_2)\partial_{2-}\phi(x_2^-,0,y_2)]\notag\\
    &=\sqrt{2}\int_{-\infty}^{\infty}dx_1^-\int_{-\infty}^{\infty}dx_2^-~x_1^-\bigg([(\partial_{1-}\phi(x_1^-,0,y_1))^2,\partial_{2-}\phi(x_2^-,0,y_2)]\psi_{\downarrow}(x_2^-,0,y_2)\notag\\
    &+\frac{i}{2}(\partial_{2-}\phi(x_2^-,0,y_2)[\psi_{\downarrow}(x_1^-,0,y_1)\partial_{1-}\psi_{\downarrow}(x_1^-,0,y_1),\psi_{\downarrow}(x_2^-,0,y_2)]\bigg)\notag\\
    &=\sqrt{2}i\delta(y_1-y_2)\int_{-\infty}^{\infty}dx_1^-\int_{-\infty}^{\infty}dx_2^-~x_1^-\bigg(\partial_{1-}\delta(x_1^--x_2^-)\partial_{1-}\phi(x_1^-,0,y_1)~\psi_{\downarrow}(x_2^-,0,y_2)\notag\\&+\frac{1}{2}\partial_{2-}\phi(x_2^-,0,y_2)\big(\psi_{\downarrow}(x_1^-,0,y_1)\partial_{1-}\delta(x_1^- -x_2^-)-\delta(x_1^--x_2^-)\partial_{1-}\psi_{\downarrow}(x_1^-,0,y_1)\big)\bigg)\notag\\
    &=\sqrt{2}i\delta(y_1-y_2)\int_{-\infty}^{\infty}dx^-\bigg(-\partial_{-}\big(x^-\partial_{-}\phi(x^-,0,y_2)\big)\psi_{\downarrow}(x^-,0,y_2)-\frac{1}{2}\partial_{-}\phi(x^-,0,y_2)\partial_{-}\big(x^- \psi_{\downarrow}(x^-,0,y_2)\big)\notag\\
    &-\frac{1}{2}x^-\partial_{-}\phi(x^-,0,y_2)\partial_{-}\psi_{\downarrow}(x^-,0,y_2)\bigg)\notag\\
    &=-\frac{i}{\sqrt{2}}\delta(y_1-y_2)\int_{-\infty}^{\infty}dx^-\phi(x^-,0,y_2)\partial_-\psi_{\downarrow}(x^-,0,y_2)=-\frac{i}{2}\delta(y_1-y_2)q(y_2),
\end{align}
\normalsize
perfectly matching our abstract derivation. Finally, we want to compute,
\begin{align}
    [\mathcal{N}_y(y_1),q(y_2)].
\end{align}
In this theory,
\begin{align}
    \mathcal{N}_y(y)=\int_{-\infty}^{\infty}dx^- T_{-y}(x^-,0,y)=\int_{-\infty}^{\infty}dx^-\bigg(\partial_-\phi\partial_y\phi+\frac{i}{2}\psi_{\downarrow}\partial_y\psi_{\downarrow}\bigg),
\end{align}
where we used the equation of motion to write $\psi_{\uparrow}$ in terms of $\psi_{\downarrow}$. We have,
\begin{align}
    [\mathcal{N}_y(y_1),q(y_2)]=\sqrt{2}\int_{-\infty}^{\infty}dx_1^-\int_{-\infty}^{\infty}dx_2^-\bigg([\partial_{1-}\phi_1\partial_{y_1}\phi_1,\psi_{2\downarrow}\partial_{2-}\phi_2]+\frac{i}{2}[\psi_{1\downarrow}\partial_{y_1}\psi_{1\downarrow},\psi_{2\downarrow}\partial_{2-}\phi_2]\bigg).
\end{align}
We evaluate these commutators using,
\begin{align}
    &[B_1 B_2,B_3]=B_1[B_2,B_3]+[B_1,B_3]B_2,\notag\\
    &[F_1F_2,F_3]=F_1\{F_2,F_3\}-\{F_1,F_3\}F_2.
\end{align}
Then, we evaluate the integral over $x_1^-$, which results in,
\begin{align}
    &[\mathcal{N}_y(y_1),q(y_2)]=\frac{i}{\sqrt{2}}\partial_{y_1}\delta(y_1-y_2)\int_{-\infty}^{\infty}dx^-\bigg(\psi_{\downarrow}(x^-,0,y_2)\partial_-\phi(x^-,0,y_1)+\psi_{\downarrow}(x^-,0,y_1)\partial_-\phi(x^-,0,y_2)\bigg)\notag\\
    &+\frac{i}{\sqrt{2}}\delta(y_1-y_2)\int_{-\infty}^{\infty}dx^-\bigg(-\psi_{\downarrow}(x^-,0,y_2)\partial_-\partial_{y_2}\phi(x^-,0,y_2)-\partial_{y_2}\psi_{\downarrow}(x^-,0,y_2)\partial_-\phi(x^-,0,y_2)\bigg).
\end{align}
One has to take care of the following distributional identity when replacing $y_1$ by $y_2$ in the integrand in the first line.
\begin{align}
    &\partial_{y_1}\big(\delta(y_1-y_2)I(y_1,y_2)\bigg)=\partial_{y_1}\bigg(\delta(y_1-y_2)I(y_2,y_2)\bigg)\notag\\
    &\implies \partial_{y_1}\delta(y_1-y_2)~I(y_1,y_2)=\partial_{y_1}\delta(y_1-y_2)~I(y_2,y_2)-\delta(y_1-y_2)\partial_{y_1}F(y_1,y_2).
\end{align}
We then find,
\begin{align}
    [\mathcal{N}_y(y_1),q(y_2)]=-i\delta(y_1-y_2)\partial_{y_2}q(y_2)+i\partial_{y_1}\delta(y_1-y_2)q(y_2),
\end{align}
which perfectly matches with our general formula.
\subsection{Four dimensions}
The action for the free massless four dimensional Wess-Zumino model is,
\begin{align}
    S=\int d^4 x\bigg(-\eta^{\mu\nu}\partial_\mu\phi\partial_\nu\phi^*+i\Bar{\psi}_{\Dot{\alpha}}(\Bar{\sigma}^\mu)^{\Dot{\alpha}\alpha}\partial_\mu\psi_{\alpha}\bigg),
\end{align}
where we have set the auxiliary field to zero using its equation of motion. The equations of motion of the complex scalar and Weyl fermion are,
\begin{align}
    &\qquad\qquad\qquad\qquad\qquad\qquad\qquad\qquad\Box\phi=0,\Box\phi^*=0,\notag\\
    &\qquad\qquad\qquad\qquad\qquad\qquad(\Bar{\sigma}^\mu)^{\Dot{\alpha}\alpha}\partial_\mu\psi_{\alpha}=0,\partial_\mu\Bar{\psi}_{\Dot{\alpha}}(\Bar{\sigma}^\mu)^{\Dot{\alpha}\alpha}=0\notag\\
    &\implies \partial_+\psi_{\downarrow}+\partial_z\psi_{\uparrow}=0,\partial_-\psi_{\uparrow}+\partial_{\Bar{z}}\psi_{\downarrow}=0,\partial_+\Bar{\psi}_{\Dot{\downarrow}}+\partial_{\Bar{z}}\Bar{\psi}_{\Dot{\uparrow}}=0,\partial_-\Bar{\psi}_{\Dot{\uparrow}}+\partial_{z}\Bar{\psi}_{\Dot{\downarrow}}=0.
\end{align}
We quantize this theory on the light-front at $x^+=0$. The propagating degrees of freedom are represented by $\phi,\phi^*,\psi_{\downarrow},\Bar{\psi}_{\downarrow}$. The canonical (anti-)commutation relations are,
\begin{align}
    &[\phi(x_1^-,0,z_1,\Bar{z}_1),\partial_{2-}\phi^*(x_2^-,0,z_2,\Bar{z}_2)]=\frac{i}{2}\delta(x_1^--x_2^-)\delta^2(z_1-z_2),\notag\\
    &\{\psi_{\downarrow}(x_1^-,0,z_1,\Bar{z}_1),\Bar{\psi}_{\Dot{\downarrow}}(x_2^-,0,z_2,\Bar{z}_2)\}=\delta(x_1^--x_2^-)\delta^2(z_1-z_2).
\end{align}
The (improved) stress tensor takes the form,
\begin{align}\label{4dfreetheoryTmunu}
    T_{\mu\nu}&=\partial_{\mu}\phi^*\partial_{\nu}\phi+\partial_{\mu}\phi\partial_\nu\phi^*-\eta_{\mu\nu}\partial_\rho\phi\partial^\rho\phi^*+\frac{1}{3}\big(\eta_{\mu\nu}\Box-\partial_\mu\partial_\nu\big)\big(\phi^*\phi\big)\notag\\
    &-\frac{i}{4}\bigg(\Bar{\psi}_{\Dot{\alpha}}(\Bar{\sigma}_\mu)^{\Dot{\alpha}\alpha}\partial_{\nu}\psi_{\alpha}-\partial_{\nu}\Bar{\psi}_{\Dot{\alpha}}(\Bar{\sigma}_\mu)^{\Dot{\alpha}\alpha}\psi_{\alpha}+\Bar{\psi}_{\Dot{\alpha}}(\Bar{\sigma}_\nu)^{\Dot{\alpha}\alpha}\partial_{\mu}\psi_{\alpha}-\partial_{\mu}\Bar{\psi}_{\Dot{\alpha}}(\Bar{\sigma}_\nu)^{\Dot{\alpha}\alpha}\psi_{\alpha}\bigg).
\end{align}
The supersymmetry currents on the other hand are,
\begin{align}\label{4dsusycurrentsfreetheory}
    &\mathcal{J}_{\mu\alpha}=2\bigg((\sigma^\nu)_{\alpha\Dot{\alpha}}(\Bar{\sigma}_{\mu})^{\Dot{\alpha}\beta}\psi_{\beta}\partial_\nu \phi^*+\frac{1}{3}\big((\sigma_{\mu})_{\alpha\Dot{\gamma}}(\Bar{\sigma}_\nu)^{\Dot{\gamma}\beta}-(\sigma_{\nu})_{\alpha\Dot{\gamma}}(\Bar{\sigma}_\mu)^{\Dot{\gamma}\beta}\big)\partial^\nu(\phi^*\psi_{\beta})\bigg),\notag\\
    &\tilde{\mathcal{J}}_{\mu\Dot{\alpha}}=2\bigg(\Bar{\psi}_{\Dot{\beta}}(\Bar{\sigma}_\mu)^{\Dot{\beta}\gamma}(\sigma^\nu)_{\gamma\Dot{\alpha}}\partial_\nu\phi+\frac{1}{3}\big((\Bar{\sigma}_{\mu})_{\Dot{\alpha}\gamma}(\sigma_\nu)^{\gamma\Dot{\beta}}-(\Bar{\sigma}_{\nu})_{\Dot{\alpha}\gamma}(\sigma_\mu)^{\gamma\Dot{\beta}}\big)\partial^\nu(\phi\Bar{\psi}_{\Dot{\beta}})\bigg).
\end{align}
Using \eqref{4dfreetheoryTmunu}, we find the averaged null energy operator,
\begin{align}
    \mathcal{E}(z,\Bar{z})=\int_{-\infty}^{\infty}dx^-\bigg(2\partial_-\phi\partial_-\phi^*+i\Bar{\psi}_{\Dot{\downarrow}}\partial_-\psi_{\downarrow}\bigg).
\end{align}
We then construct the supersymmetric light-ray operators using \eqref{4dsusycurrentsfreetheory}.
\begin{align}
    &q(z,\Bar{z})=2\int_{-\infty}^{\infty}dx^-\psi_{\downarrow}\partial_-\phi^*,\notag\\
    &\tilde{q}(z,\Bar{z})=2\int_{-\infty}^{\infty}dx^-\Bar{\psi}_{\Dot{\downarrow}}\partial_-\phi,
\end{align}
where all fields are normal ordered and are located at $(x^-,x^+=0,z,\Bar{z})$ in the integrand. The $\mathcal{K}$ and $\mathcal{N}_i$ operators take the form,
\begin{align}
    &\mathcal{K}(z,\Bar{z})=\int_{-\infty}^{\infty}dx^-x^-\bigg(2\partial_-\phi\partial_-\phi^*+\frac{i}{2}(\Bar{\psi}_{\Dot{\downarrow}}\partial_-\psi_{\downarrow}-\partial_-\Bar{\psi}_{\Dot{\downarrow}}\psi_{\downarrow})\bigg),\notag\\
    &\mathcal{N}_i(z,\Bar{z})=\int_{-\infty}^{\infty}dx^-\Bigg(\partial_-\phi^*\partial_i\phi+\partial_-\phi\partial_i\phi^*-\frac{i}{4}(\partial_i\Bar{\psi}_{\Dot{\downarrow}}~\psi_{\downarrow}-\Bar{\psi}_{\Dot{\downarrow}}\partial_i\psi_{\downarrow})+\frac{i}{2}\big(\delta_i^z\psi_{\downarrow}\partial_z \Bar{\psi}_{\Dot{\downarrow}}+\delta_i^{\Bar{z}}\Bar{\psi}_{\Dot{\downarrow}}\partial_{\Bar{z}}\psi_{\downarrow})\bigg),
\end{align}
where $i\in\{z,\Bar{z}\}$. We now compute $\{q,\tilde{q}\}, \{\mathcal{K},q\},\{\mathcal{K},\tilde{q}\}$, $\{\mathcal{N}_i,q\}$ and $\{\mathcal{N}_i,\tilde{q}\}$ using the free theory form of these operators and the canonical (anti-)commutation relations. We also make use of the following identities. Let $F_1,F_2$ be fermionic operators and $B_1,B_2$ be bosonic. Then, \begin{align}
    \{F_1 B_1,F_2 B_2\}=\{F_1,F_2\}B_1 B_2-F_2 F_1[B_1,B_2]
\end{align}

The anti-commutator of the supersymmetric light-ray operator with its Hermitian conjugate is,
\begin{align}
    &\{q(z_1,\Bar{z}_1),\tilde{q}(z_2,\Bar{z}_2)\}=4\int_{-\infty}^{\infty}dx_1^-\int_{-\infty}^{\infty}dx_2^-\{\psi_{1\downarrow}\partial_{1-}\phi_1^*,\Bar{\psi}_{2\downarrow}\partial_{2-}\phi_2)\}\notag\\&=4\int_{-\infty}^{\infty}dx_1^-\int_{-\infty}^{\infty}dx_2^-\bigg(\delta(x_1^--x_2^-)\delta^2(z_{12})\partial_{1-}\phi_1^*\partial_{2-}\phi_2-\frac{i}{2}\Bar{\psi}_{2\downarrow}\psi_{1\downarrow}\delta^2(z_{12})\partial_{1-}\delta(x_1^-x_2^-)\bigg)\notag\\
    &=2\delta^2(z_{12})\int_{-\infty}^{\infty}dx^-\bigg(2\partial_-\phi^*\partial_-\phi+i\Bar{\psi}_{\downarrow}\partial_-\psi_{\downarrow}\bigg)=2\delta^2(z_{12})\mathcal{E}(z_2,\Bar{z}_2).
\end{align}
Next, we compute the commutators of $q$ and $\tilde{q}$ with $\mathcal{K}$.
\begin{align}
    &[\mathcal{K}(z_1,\Bar{z}_1),q(z_2,\Bar{z}_2)]=2\int_{-\infty}^{\infty}dx_1^- x_1^-\int_{-\infty}^{\infty}dx_2^-\bigg(2[\partial_{1-}\phi_1\partial_{1-}\phi_1^*,\psi_{2\downarrow}\partial_{2-}\phi_2^*]\notag\\
    &+\frac{i}{2}[\Bar{\psi}_{1\downarrow}\partial_{1-}\psi_{1\downarrow},\psi_{2\downarrow}\partial_{2-}\phi_2^*]+\frac{i}{2}[\psi_{1\downarrow}\partial_{1-}\Bar{\psi}_{1\downarrow},\psi_{2\downarrow}\partial_{2-}\phi_2^*]\bigg)\notag\\
    &=2\delta^2(z_{12})\int_{-\infty}^{\infty}dx_1^- x_1^-\int_{-\infty}^{\infty}dx_2^-\bigg(i\partial_{1-}\phi_1^*\psi_{2\downarrow}\partial_{1-}\delta(x_1^--x_2^-)-\frac{i}{2}\partial_{2-}\phi_2^*\delta(x_1^--x_2^-)\partial_{1-}\psi_{1\downarrow}\notag\\
    &+\frac{i}{2}\partial_{2-}\phi_2^*\psi_{1\downarrow}\partial_{1-}\delta(x_1^--x_2^-)\bigg)\notag\\
    &=2\delta^2(z_{12})\int_{-\infty}^{\infty}dx^-\bigg(-i\psi_{\downarrow}\partial_-\phi^*-i x^-\psi_{\downarrow}\partial_-^2\phi^*-\frac{ix^-}{2}\partial_{-}\psi_{\downarrow}\partial_{-}\phi^*-\frac{i}{2}x^-\partial_-\phi^*\partial_-\psi_{\downarrow}-\frac{i}{2}\psi_{\downarrow}\partial_-\phi^*\bigg)\notag\\
    &=-\frac{i}{2}\delta^2(z_{12})q(z_2,\Bar{z}_2),
\end{align}
where in going from the second step to the third, we performed the integration by parts and computed the $x_1^-$ integral. The commutator with $\tilde{q}$ is obtained in a similar way.
\begin{align}
    &[\mathcal{K}(z_1,\Bar{z}_1),\tilde{q}(z_2,\Bar{z}_2)]=2\int_{-\infty}^{\infty}dx_1^- x_1^-\int_{-\infty}^{\infty}dx_2^-\bigg(2[\partial_{1-}\phi_1\partial_{1-}\phi_1^*,\Bar{\psi}_{2\downarrow}\partial_{2-}\phi_2]\notag\\
    &+\frac{i}{2}[\Bar{\psi}_{1\downarrow}\partial_{1-}\psi_{1\downarrow},\Bar{\psi}_{2\downarrow}\partial_{2-}\phi_2]+\frac{i}{2}[\psi_{1\downarrow}\partial_{1-}\Bar{\psi}_{1\downarrow},\Bar{\psi}_{2\downarrow}\partial_{2-}\phi_2]\bigg)\notag\\
    &=2\delta^2(z_{12})\int_{-\infty}^{\infty}dx_1^- x_1^-\int_{-\infty}^{\infty}dx_2^-\bigg(i\partial_{1-}\phi_1\Bar{\psi}_{2\downarrow}\partial_{1-}\delta(x_1^--x_2^-)+\frac{i}{2}\partial_{2-}\phi_2\partial_{1-}\delta(x_1^--x_2^-)\Bar{\psi}_{1\downarrow}\notag\\
    &+\frac{i}{2}\partial_{2-}\phi_2\partial_{1-}\Bar{\psi}_{1\downarrow}\delta(x_1^--x_2^-)\bigg)\notag\\
    &=2\delta^2(z_{12})\int_{-\infty}^{\infty}dx^-\bigg(-ix^-\partial_-^2\phi\Bar{\psi}_{\downarrow}-i\partial_-\phi\Bar{\psi}_{\downarrow}-\frac{i}{2}\Bar{\psi}_{\downarrow}\partial_-\phi-ix^-\partial_-\Bar{\psi}_{\downarrow}\partial_-\phi\bigg)\notag\\
    &=-\frac{i}{2}\delta^2(z_{12})\tilde{q}(z_2,\Bar{z}_2),
\end{align}
where we performed an integration by parts to arrive at the result. Finally, we compute the most non-trivial commutators, that being, the ones with $\mathcal{N}_i$. First we take $i=z$ and compute,
\begin{align}
    &[\mathcal{N}_z(z_1,\Bar{z}_1),q(z_2,\Bar{z}_2)]
    =2\int_{-\infty}^{\infty}dx_1^-
      \int_{-\infty}^{\infty}dx_2^-\bigg(
      [\partial_{1-}\phi_1^*\,\partial_{z_1}\phi_1,
        \psi_{2\downarrow}\partial_{2-}\phi_2^*]
      +[\partial_{1-}\phi_1\,\partial_{z_1}\phi_1^*,
        \psi_{2\downarrow}\partial_{2-}\phi_2^*]\notag\\
    &\hspace{35mm}
      -\frac{i}{4}[\partial_{z_1}\Bar{\psi}_{1\downarrow}
        \psi_{1\downarrow},
        \psi_{2\downarrow}\partial_{2-}\phi_2^*]
      -\frac{i}{4}[\partial_{z_1}\psi_{1\downarrow}
        \Bar{\psi}_{1\downarrow},
        \psi_{2\downarrow}\partial_{2-}\phi_2^*]
      +\frac{i}{2}[\psi_{1\downarrow}
        \partial_{z_1}\Bar{\psi}_{1\downarrow},
        \psi_{2\downarrow}\partial_{2-}\phi_2^*]\bigg)\notag\\
    &=2\int_{-\infty}^{\infty}dx_1^-
      \int_{-\infty}^{\infty}dx_2^-\bigg(
      \frac{i}{2}\partial_{1-}\phi_1^*
        \psi_{2\downarrow}\partial_{z_1}\delta^2(z_{12})
        \delta(x_1^--x_2^-)\notag\\
    &\hspace{35mm}
      +\frac{i}{2}\psi_{2\downarrow}\partial_{z_1}\phi_1^*
        \delta^2(z_{12})\partial_{1-}\delta(x_1^--x_2^-)
      +\frac{i}{4}\partial_{z_1}\delta^2(z_{12})
        \delta(x_1^--x_2^-)\psi_{1\downarrow}
        \partial_{2-}\phi_2^*\notag\\
    &\hspace{35mm}
      -\frac{i}{4}\partial_{z_1}\psi_{1\downarrow}
        \partial_{2-}\phi_2^*\delta^2(z_{12})
        \delta(x_1^--x_2^-)
      +\frac{i}{2}\psi_{1\downarrow}
        \partial_{2-}\phi_2^*\partial_{z_1}\delta^2(z_{12})
        \delta(x_1^--x_2^-)\bigg)\notag\\
    &=i\partial_{z_1}\delta^2(z_{12})
      \int_{-\infty}^{\infty}dx^-\bigg(
        \partial_-\phi^*(x^-,0,z_1)\psi_\downarrow(x^-,0,z_2)
        +\frac{3}{2}\psi_\downarrow(x^-,0,z_1)
          \partial_-\phi^*(x^-,0,z_2)\bigg)\notag\\
    &\quad-i\delta^2(z_{12})
      \int_{-\infty}^{\infty}dx^-\bigg(
        \psi_\downarrow(x^-,0,z_2)
          \partial_{z_1}\partial_-\phi^*(x^-,0,z_1)
        +\frac{1}{2}\partial_{z_1}\psi_\downarrow(x^-,0,z_1)
          \partial_-\phi^*(x^-,0,z_2)\bigg).
\end{align}
To proceed, we make use of the following distributional identity:
\begin{align}
    \partial_{z_1}\delta^2(z_{12})f(z_1)g(z_2)
    =\partial_{z_1}\delta^2(z_{12})f(z_2)g(z_2)
     -\delta^2(z_{12})\partial_{z_2}f(z_2)g(z_2).
\end{align}
This results in,
\begin{align}
    &[\mathcal{N}_z(z_1,\Bar{z}_1),q(z_2,\Bar{z}_2)]\notag\\
    &=i\partial_{z_1}\delta^2(z_{12})
      \int_{-\infty}^{\infty}dx^-\bigg(
        \partial_-\phi^*(x^-,0,z_2)\psi_\downarrow(x^-,0,z_2)
        +\frac{3}{2}\psi_\downarrow(x^-,0,z_2)
          \partial_-\phi^*(x^-,0,z_2)\bigg)\notag\\
    &\quad-i\delta^2(z_{12})
      \int_{-\infty}^{\infty}dx^-\bigg(
        \psi_\downarrow(x^-,0,z_2)
          \partial_{z_2}\partial_-\phi^*(x^-,0,z_2)
        +\frac{1}{2}\partial_{z_2}\psi_\downarrow(x^-,0,z_2)
          \partial_-\phi^*(x^-,0,z_2)\notag\\
    &\hspace{47mm}
        +\partial_{z_2}\partial_-\phi^*(x^-,0,z_2)
          \psi_\downarrow(x^-,0,z_2)
        +\frac{3}{2}\partial_{z_2}\psi_\downarrow(x^-,0,z_2)
          \partial_-\phi^*(x^-,0,z_2)\bigg)\notag\\
    &=\frac{5i}{4}\partial_{z_1}\delta^2(z_{12})
        q(z_2,\Bar{z}_2)
      -i\delta^2(z_{12})\partial_{z_2}q(z_2,\Bar{z}_2)\notag\\
    &=-i\delta^2(z_{12})\partial_{z_2}q(z_2,\Bar{z}_2)
      +i\partial_{z_1}\delta^2(z_{12})q(z_2,\Bar{z}_2)
      +\frac{i}{4}\partial_{z_1}\delta^2(z_{12})
        q(z_2,\Bar{z}_2).
\end{align}
Next, we take $i=\Bar{z}$ and compute,
\begin{align}
    &[\mathcal{N}_{\Bar{z}}(z_1,\Bar{z}_1),q(z_2,\Bar{z}_2)]
    =2\int_{-\infty}^{\infty}dx_1^-
      \int_{-\infty}^{\infty}dx_2^-\bigg(
      [\partial_{1-}\phi_1^*\,\partial_{\Bar{z}_1}\phi_1,
        \psi_{2\downarrow}\partial_{2-}\phi_2^*]
      +[\partial_{1-}\phi_1\,\partial_{\Bar{z}_1}\phi_1^*,
        \psi_{2\downarrow}\partial_{2-}\phi_2^*]\notag\\
    &\hspace{35mm}
      -\frac{i}{4}[\partial_{\Bar{z}_1}\Bar{\psi}_{1\downarrow}
        \psi_{1\downarrow},
        \psi_{2\downarrow}\partial_{2-}\phi_2^*]
      -\frac{i}{4}[\partial_{\Bar{z}_1}\psi_{1\downarrow}
        \Bar{\psi}_{1\downarrow},
        \psi_{2\downarrow}\partial_{2-}\phi_2^*]
      +\frac{i}{2}[\Bar{\psi}_{1\downarrow}
        \partial_{\Bar{z}_1}\psi_{1\downarrow},
        \psi_{2\downarrow}\partial_{2-}\phi_2^*]\bigg)\notag\\
    &=2\int_{-\infty}^{\infty}dx_1^-
      \int_{-\infty}^{\infty}dx_2^-\bigg(
      \frac{i}{2}\partial_{1-}\phi_1^*\psi_{2\downarrow}
        \partial_{\Bar{z}_1}\delta^2(z_{12})
        \delta(x_1^--x_2^-)\notag\\
    &\hspace{35mm}
      +\frac{i}{2}\psi_{2\downarrow}
        \partial_{\Bar{z}_1}\phi_1^*
        \delta^2(z_{12})\partial_{1-}\delta(x_1^--x_2^-)
      +\frac{i}{4}\partial_{\Bar{z}_1}\delta^2(z_{12})
        \delta(x_1^--x_2^-)\psi_{1\downarrow}
        \partial_{2-}\phi_2^*\notag\\
    &\hspace{35mm}
      -\frac{i}{4}\partial_{\Bar{z}_1}\psi_{1\downarrow}
        \partial_{2-}\phi_2^*\delta^2(z_{12})
        \delta(x_1^--x_2^-)
      -\frac{i}{2}\partial_{\Bar{z}_1}\psi_{1\downarrow}
        \partial_{2-}\phi_2^*\delta^2(z_{12})
        \delta(x_1^--x_2^-)\bigg)\notag\\
    &=i\partial_{\Bar{z}_1}\delta^2(z_{12})
      \int_{-\infty}^{\infty}dx^-\bigg(
        \partial_-\phi^*(x^-,0,z_1)\psi_\downarrow(x^-,0,z_2)
        +\frac{1}{2}\psi_\downarrow(x^-,0,z_1)
          \partial_-\phi^*(x^-,0,z_2)\bigg)\notag\\
    &\quad-i\delta^2(z_{12})
      \int_{-\infty}^{\infty}dx^-\bigg(
        \psi_\downarrow(x^-,0,z_2)
          \partial_{\Bar{z}_1}\partial_-\phi^*(x^-,0,z_1)
        +\frac{3}{2}\partial_{\Bar{z}_1}
          \psi_\downarrow(x^-,0,z_1)
          \partial_-\phi^*(x^-,0,z_2)\bigg)\notag\\
    &=\frac{3i}{4}\partial_{\Bar{z}_1}\delta^2(z_{12})
        q(z_2,\Bar{z}_2)
      -i\delta^2(z_{12})
        \partial_{\Bar{z}_2}q(z_2,\Bar{z}_2)\notag\\
    &=-i\delta^2(z_{12})
        \partial_{\Bar{z}_2}q(z_2,\Bar{z}_2)
      +i\partial_{\Bar{z}_1}\delta^2(z_{12})
        q(z_2,\Bar{z}_2)
      -\frac{i}{4}\partial_{\Bar{z}_1}\delta^2(z_{12})
        q(z_2,\Bar{z}_2).
\end{align}
Thus, we have found in the free theory that,
\begin{align}
    [\mathcal{N}_z(z_1,\Bar{z}_1),q(z_2,\Bar{z}_2)]
    &=-i\delta^2(z_{12})\partial_{z_2}q(z_2,\Bar{z}_2)
      +i\partial_{z_1}\delta^2(z_{12})q(z_2,\Bar{z}_2)+\frac{i}{4}\partial_{z_1}\delta^2(z_{12})
      q(z_2,\Bar{z}_2),\notag\\[4pt]
    [\mathcal{N}_{\Bar{z}}(z_1,\Bar{z}_1),q(z_2,\Bar{z}_2)]
    &=-i\delta^2(z_{12})\partial_{\Bar{z}_2}q(z_2,\Bar{z}_2)
      +i\partial_{\Bar{z}_1}\delta^2(z_{12})
      q(z_2,\Bar{z}_2)-\frac{i}{4}\partial_{\Bar{z}_1}\delta^2(z_{12})
      q(z_2,\Bar{z}_2).
\end{align}
Combining these relations, we obtain,
\begin{align}
    [\mathcal{N}_i(z_1,\Bar{z}_1),q(z_2,\Bar{z}_2)]
    &=-i\delta^2(z_{12})\partial_{i2}q(z_2,\Bar{z}_2)
      +i\partial_{i1}\delta^2(z_{12})q(z_2,\Bar{z}_2)+\frac{1}{4}\epsilon_{i}^{j}\partial_{j1}
      \delta^2(z_{12})q(z_2,\Bar{z}_2),
\end{align}
where $\epsilon_z^z=i, \epsilon_{\Bar{z}}^{\Bar{z}}=-i$. This is exactly the result we presented in the main-text for general theories. A similar calculation then shows that,
\begin{align}
    [\mathcal{N}_i(z_1,\Bar{z}_1),\tilde{q}(z_2,\Bar{z}_2)]
    &=-i\delta^2(z_{12})\partial_{i2}\tilde{q}(z_2,\Bar{z}_2)
      +i\partial_{i1}\delta^2(z_{12})\tilde{q}(z_2,\Bar{z}_2)-\frac{1}{4}\epsilon_{i}^{j}\partial_{j1}
      \delta^2(z_{12})\tilde{q}(z_2,\Bar{z}_2),
\end{align}
which also matches with our general result. Thus, we have verified our algebra in the free theory.

\section{Generalization of the ANEC proof to supersymmetric QFT}\label{app:QFT}
In this appendix, we discuss the extension of the supersymmetric light-ray algebra to non-conformal supersymmetric quantum field theories. We focus on $d=3$ for concreteness. What is true in every supersymmetric quantum field theory is,
\begin{align}
    \{Q_{\downarrow},q(y)\}=2\mathcal{E}(y).
\end{align}
This just follows from the action of the global supercharge on the super-current.
\begin{align}
    \{Q_{\downarrow},q(y)\}=\int_{-\infty}^{\infty}dx^-\{Q_{\downarrow},\mathcal{J}_{-\downarrow}(x^-,0,y)\}=2\int_{-\infty}^{\infty}dx^- T_{--}(x^-,0,y)=2\mathcal{E}(y).
\end{align}
Clearly, this statement is not enough to prove the ANEC. To see this, let us smear both sides of this equation and take the expectation value in a state $|\Psi\rangle$. We find,
\begin{align}
    &\int_{-\infty}^{\infty}dy f(y)\langle \Psi|\{Q_{\downarrow},q(y)\}|\Psi\rangle=2\int_{-\infty}^{\infty}dy f(y) \langle \Psi|\mathcal{E}(y)|\Psi\rangle,
\end{align}
which does not generally imply the positivity of the operator $\mathcal{E}(y)$. It is the local algebra we have derived,
\begin{align}
   &\{q(y_1),q(y_2)\}=2\delta(y_1-y_2)\mathcal{E}(y_2),\notag\\
   &\implies \{\mathcal{Q}(f),\mathcal{Q}(f)\}=2\mathcal{T}(f^2),
\end{align}
that ensures the ANEC holds. Let us return to the local algebra ansatz,
\begin{align}
    \{q(y_1),q(y_2)\}=2\delta(y_1-y_2)\mathcal{E}(y_2)+\sum_{n=1}^{\infty}\partial_{y_1}^{n}\delta(y_1-y_2)A_n(y_2).
\end{align}
In the main-text we have shown that imposing conformal symmetry and unitarity sets the $A_n,n\ge 1$ to zero. In a non-conformal supersymmetric quantum field theory these terms could potentially be generated and  contribute. We now present an argument similar to the one in \cite{Casini:2017roe} to show that under certain assumptions, the algebra remains undeformed even beyond conformality. 
\subsection{A General Argument}
Consider a mass deformation of the UV supersymmetric conformal field theory $SCFT$ that preserves the supersymmetry.
\begin{align}
    S_{SQFT}=S_{SCFT}+g\int d^3 x~\mathcal{O}(x).
\end{align}
 $\mathcal{O}$ has scaling dimension $\Delta$ and $g$ has mass dimension $r=3-\Delta$. We consider relevant deformations so $\Delta<3$ and as a consequence, $r>0$. The deformed theory could flow to another conformal fixed point or could be massive. Consider the potentially deformed light-ray supersymmetry algebra at some perturbative order:
\begin{align}
    \{q(y_1),q(y_2)\}=2\delta(y_1-y_2)\mathcal{E}(y_2)+\sum_{i=1}^{\infty}\sum_{n=1}^{\infty}g^i\partial_{y_1}^{n}\delta(y_1-y_2)L_{n,i}(y).
\end{align}
At each order in perturbation theory, we are allowing for the possible generation of arbitrary derivatives of $\delta(y_1-y_2)$. Let us now count the boost-eigen value on both sides of this equation. The left hand side has $m=1$ as does the first term on the right hand side. This also implies that $L_{n,i}(y)$ must have boost weight $1$. We now assume closure just like we did in the main-text writing $L_{n,i}$ as the null integral of a local operator,
\begin{align}
    L_{n,i}(y)=\int_{-\infty}^{\infty}dx^- A_{n,i}(x^-,0,y).
\end{align}
Thus $A_{n,m}$ must have boost eigen-value $2$. As for the mass dimension, the left hand side has total mass dimension $3$. Thus we find,
\begin{align}
    i r+n+1+\Delta_{L_{n,i}}=3\implies \Delta_{L_{n,i}}=2-n-i r.
\end{align}
Thus,
\begin{align}
    \Delta_{A_{n,i}}=3-n-i r.
\end{align}
Since $A_{n,i}$ must have boost weight equal to $2$, it must be constructed out of a UV operator with spin $s\ge 2$. Imposing unitarity yields,
\begin{align}
    \Delta_{A_{n,i}}\ge m_{A_{n,i}}+1\implies 3-n-ir\ge 3\implies n+i r\le 0.
\end{align}
However, $n>0$ and $i>0$ so this inequality has no non-trivial solutions. Thus, all the $A_{n,i}$, $n\ge 1,i\ge 1$ are identically zero, yielding the undeformed algebra,
\begin{align}
    \{q(y_1),q(y_2)\}=2\delta(y_1-y_2)\mathcal{E}(y_2),
\end{align}
in supersymmetric quantum field theories (not necessarily conformal). Therefore, given this equation, our supersymmetric proof of the null energy condition in section \ref{sec:ANEC} is extended to supersymmetric quantum field theories. We leave a more rigorous check for future work.
\subsection{Testing the hypothesis in the massive interacting Wess-Zumino Theory}
Consider the three dimensional $\mathcal{N}=1$ Wess-Zumino action with a general super-potential,
\begin{align}
    S=\int d^3 x~\bigg(-\frac{1}{2}\eta^{\mu\nu}\partial_\mu\phi\partial_\nu \phi-\frac{i}{2}\psi^a \slashed{\partial}_{ab}\psi^b-\frac{1}{2}W'(\phi)^2+\frac{i}{2}W''(\phi)\psi_a\psi^a\bigg).
\end{align}
For example let us take,
\begin{align}
    W(\phi)=\frac{m}{2}\phi^2+\frac{g}{3}\phi^3.
\end{align}
This results in,
\begin{align}
    S=\int d^3 x~\bigg(-\frac{1}{2}\eta^{\mu\nu}\partial_\mu\phi\partial_\nu \phi-\frac{i}{2}\psi^a \slashed{\partial}_{ab}\psi^b-\frac{1}{2}m^2\phi^2+\frac{i}{2}m\psi_a\psi^a-\frac{g^2}{8}\phi^4-\frac{mg}{2}\phi^3+\frac{ig}{2}\phi\psi_a\psi^a\bigg).
\end{align} 
The corrections to the stress tensor and supersymmetry current are as follows (for a general super-potential)
\begin{align}
    T_{\mu\nu}&=\bigg(\partial_\mu\phi\partial_\nu\phi-\frac{\eta_{\mu\nu}}{2}(\partial \phi)^2+\frac{1}{8}(\eta_{\mu\nu}\Box-\partial_\mu\partial_\nu)(\phi^2)\bigg)\notag\\
    &-\frac{i}{8}\bigg(\psi^a(\gamma_\mu)_{ab}\partial_\nu \psi^b-\partial_\nu\psi^a(\gamma_\mu)_{ab}\psi^b+\psi^a(\gamma_\nu)_{ab}\partial_\mu \psi^b-\partial_\mu\psi^a(\gamma_\nu)_{ab}\psi^b\bigg)\notag\\
    &-\frac{\eta_{\mu\nu}}{2}(W'(\phi))^2
\end{align}
Therefore, we see that the $--$ component does not receive contributions from the potential terms and thus remains unchanged.
\begin{align}
    T_{--}=(\partial_-\phi)^2+\frac{i}{2}\psi_{\downarrow}\partial_-\psi_{\downarrow}.
\end{align}
The form of the supersymmetry current is modified,
\begin{align}
    \mathcal{J}_{abc}&=\frac{1}{2\sqrt{2}}\bigg((\psi_a\partial_{bc}\phi+\psi_b\partial_{ca}\phi+\psi_c\partial_{ab}\phi)-\frac{1}{3}\big(\partial_{bc}\psi_a~\phi+\partial_{ca}\psi_b~\phi+\partial_{ab}\psi_c~\phi\big)\bigg)\notag\\
    &-\sqrt{2}\big(W'(\phi)-\frac{\phi}{3}W''(\phi)\big)\epsilon_{a(b}\psi_c).
\end{align}
However, its null component remains the same since $\epsilon_{\downarrow\downarrow}=0$ thus yielding,
\begin{align}
    \mathcal{J}_{\downarrow\downarrow\downarrow}=\frac{3}{2\sqrt{2}}\bigg(\psi_{\downarrow}\partial_{\downarrow\downarrow}\phi-\frac{1}{3}\partial_{\downarrow\downarrow}\psi_{\downarrow}~\phi\bigg).
\end{align}
Therefore, $\mathcal{E}(y)$ and $q(y)$ take the same form as in the massless free theory and thus the light-ray supersymmetry algebra remains the same. Of course, this analysis does not take into account the renormalization of these composite operators due to UV divergences. However, correlators involving $T_{\mu\nu}$ and $\mathcal{J}_{abc}$ must satisfy the appropriate Ward-Takahashi identities and they are responsible for generating the global translation and supersymmetry transformations. Thus, it could be possible to argue that the light-ray supersymmetry algebra is undeformed beyond conformality. We leave a more detailed analysis for the future.
\section{The Stabilizer subgroup}\label{app:PoincareFromBMS}
In this appendix, we discuss the stabilizer subgroup of the null hypersurface $x^+=0$. This is a subgroup of the super-conformal group that is isomorphic to the super-Poincare group. Consider the supersymmetric $\mathfrak{bms}_3$ generators $\mathcal{T}(f),\mathcal{R}(Y)$ and $\mathcal{Q}(g)$. We focus on $\mathcal{N}=1$ theories in three dimensions for concreteness but it is a simple matter to generalize the discussion to higher supersymmetry by including the $R-$symmetry generators as well as to $d=4$. The conformal generators in three dimensions are given by,
\begin{align}
P_\mu
&=\int_{-\infty}^{\infty}dx^-\int_{-\infty}^{\infty}dy\,
T_{-\mu}(x^-,0,y),\notag\\
M_{\mu\nu}
&=\int_{-\infty}^{\infty}dx^-\int_{-\infty}^{\infty}dy\,
\bigl(x_\mu T_{-\nu}-x_\nu T_{-\mu}\bigr)(x^-,0,y),\notag\\
D
&=\int_{-\infty}^{\infty}dx^-\int_{-\infty}^{\infty}dy\,
x^\mu T_{-\mu}(x^-,0,y),\notag\\
K_\mu
&=\int_{-\infty}^{\infty}dx^-\int_{-\infty}^{\infty}dy\,
\bigl(-2x_\mu x^\nu+x^2\delta_\mu{}^\nu\bigr)
T_{-\nu}(x^-,0,y),\notag\\
Q_a
&=\int_{-\infty}^{\infty}dx^-\int_{-\infty}^{\infty}dy\,
\mathcal{J}_{-a}(x^-,0,y),\notag\\
S_a
&=\int_{-\infty}^{\infty}dx^-\int_{-\infty}^{\infty}dy\,
x_\mu(\gamma^\mu)_a^b \mathcal{J}_{-b}(x^-,0,y).
\end{align}
It is then easy to see that,
\begin{align}
\mathcal{T}(1)
&=\int_{-\infty}^{\infty}dy\int_{-\infty}^{\infty}dx^-\,
T_{--}(x^-,0,y)=P_-,\notag\\
\mathcal{T}(y)
&=\int_{-\infty}^{\infty}dy\int_{-\infty}^{\infty}dx^-\,
yT_{--}(x^-,0,y)=M_{y-},\notag\\
\mathcal{T}(y^2)
&=\int_{-\infty}^{\infty}dy\int_{-\infty}^{\infty}dx^-\,
y^2T_{--}(x^-,0,y)=K_-,\notag\\
\mathcal{R}(1)
&=\int_{-\infty}^{\infty}dy\int_{-\infty}^{\infty}dx^-\,
T_{-y}(x^-,0,y)=P_y,\notag\\
\mathcal{R}(y)
&=\int_{-\infty}^{\infty}dy\int_{-\infty}^{\infty}dx^-\,
\bigl(yT_{-y}(x^-,0,y)+x^-T_{--}(x^-,0,y)\bigr)=D,\notag\\
\mathcal{R}(y^2)
&=\int_{-\infty}^{\infty}dy\int_{-\infty}^{\infty}dx^-\,
\bigl(y^2T_{-y}(x^-,0,y)+2yx^-T_{--}(x^-,0,y)\bigr)=-K_y,\notag\\
\mathcal{Q}(1)
&=\int_{-\infty}^{\infty}dy\int_{-\infty}^{\infty}dx^-\,
J_{-\downarrow}(x^-,0,y)=Q_\downarrow,\notag\\
\mathcal{Q}(y)
&=\int_{-\infty}^{\infty}dy\int_{-\infty}^{\infty}dx^-\,
yJ_{-\downarrow}(x^-,0,y)=S_\downarrow.
\end{align}
The $\mathcal{T}(y^n),n=0,1,2$ all mutually commute and act as translations. $\mathcal{R}(y^n),n=0,1,2$ act as rotations and thus together with $\mathcal{T}(y^n),n=0,1,2$, form a global Poincare algebra $ISO(2,1)$. Including $\mathcal{Q}(1)$ and $\mathcal{Q}(y)$ then yields the $\mathcal{N}=1$ supersymmetric Poincare algebra. 

\section{Justification of the Closure Hypothesis}\label{app:commutators}
In the main-text, we made an assumption following \cite{Cordova:2018ygx} that the commutator of light-ray operators constructed out of conserved currents only results in light-ray operators constructed out of local operators. In this appendix, we justify this assumption using the results of \cite{Besken:2020snx} in tandem with the super-conformal algebra and the graded Jacobi identities. This also provides an alternate derivation of our results. In \cite{Besken:2020snx}, the authors prove the C{\'o}rdova Shao algebra by deriving the local commutator of stress tensors on the null sheet $x^+=0$, justifying the closure assumption. We now use the global super-conformal algebra to derive from their results, the algebra involving our supersymmetric light-ray operator. We perform the analysis in $\mathcal{N}=1$ theories in $d=3$ since the generalization to higher supersymmetry as well as to $d=4$ is similar, albeit, technically more complicated\footnote{The authors of \cite{Besken:2020snx} derive the C{\'o}rdova-Shao algebra in $d=4$ but the same techniques and methods should 
generalize to $d=3$.}.
We make use of the supersymmetry and special super-conformal generators $Q_{\downarrow}$ and $S_{\downarrow}$. As we saw in appendix \ref{app:PoincareFromBMS}, these quantities are given by,
\begin{align}
    &Q_{\downarrow}=\int_{-\infty}^{\infty}dy~q(y),\notag\\
    &S_{\downarrow}=\int_{-\infty}^{\infty}dy~y~q(y).
\end{align}
Supersymmetry implies that,
\begin{align}
    &[Q_{\downarrow},\mathcal{J}_{-\downarrow}(x^-,0,y)]=2T_{--}(x^-,0,y),\notag\\
    &[Q_{\downarrow},T_{--}(x^-,0,y)]=-\frac{i}{2}\partial_-\mathcal{J}_{-\downarrow}(x^-,0,y).
\end{align}
From the global super-conformal algebra, we know that,
\begin{align}
    &[P_{\mu},S_a]=i(\sigma_\mu)_a^b Q_b\implies [P_{-},S_{\downarrow}]=0,[P_+,S_{\downarrow}]=iQ_{\uparrow},[P_y,S_{\downarrow}]=iQ_{\downarrow}.
\end{align}
The supersymmetry current is a super-conformal primary and thus it is annihilated by $S_{\downarrow}$ at the origin.
\begin{align}
    \{S_\downarrow,\mathcal{J}_{-\downarrow}(0,0,0)\}=0.
\end{align}
We also find that,
\begin{align}
    &[S_\downarrow,T_{--}(0)]=\frac{1}{2}[S_{\downarrow},\{Q_{\downarrow},\mathcal{J}_{-\downarrow}(0,0,0)\}]=\frac{1}{2}[\{S_{\downarrow},Q_{\downarrow}\},\mathcal{J}_{-\downarrow}(0,0,0)]-[Q_{\downarrow},\{S_{\downarrow},\mathcal{J}_{-\downarrow}(0,0,0\}]=0,
\end{align}
since the second term vanishes and the first also does once we use $\{S_\downarrow,Q_\downarrow\}\sim \epsilon_{\downarrow\downarrow}D+M_{\downarrow\downarrow}$, $[M_{ab},\mathcal{J}_{cde}(0,0,0)]\sim \epsilon_{ac}\mathcal{J}_{bde}+\cdots$ which vanishes when all spinor indices are identical. We now determine the action of $S_{\downarrow}$ away from the origin. We find,
\begin{align}
    [S_{\downarrow},T_{--}(x^-,x^+,y)]=U[U^{-1}S_{\downarrow}U,T_{--}(0,0,0)]U^{-1},
\end{align}
where $U=e^{i x^-P_-+ix^+ P_++i y P_y}$ is the translation generator. We first find using the Baker-Campbell-Hausdorff expansion,
\begin{align}
    U^{-1}S_{\downarrow}U=S_{\downarrow}-i x^-[P_-,S_{\downarrow}]-i x^+[P_+,S_{\downarrow}]-i y [P_y,S_{\downarrow}]=S_{\downarrow}+x^{+}Q_{\downarrow}+y Q_{\downarrow},
\end{align}
where the remaining terms dropped out since the supersymmetry generators and translations commute. Therefore,
\begin{align}
     &[S_{\downarrow},T_{--}(x^-,x^+,y)]=x^{+}[Q_{\downarrow},T_{--}(x^-,x^{+},y)]+y[Q_{\downarrow},T_{--}(x^-,x^{+},y)]\notag\\
     &\implies [S_{\downarrow},T_{--}(x^-,0,y)]=y[Q_{\downarrow},T_{--}(x^-,0,y)]=-\frac{i y}{2}\partial_-\mathcal{J}_{-\downarrow}(x^-,0,y).
\end{align}
Let us now use the actions of $Q_{\downarrow}$ and $S_{\downarrow}$ on the stress tensor and supersymmetry current to derive our supersymmetric C{\'o}rdova Shao algebra. Consider,
\begin{align}
    [Q_{\downarrow},[\mathcal{K}(y_1),\mathcal{K}(y_2)]]=0=[[Q_{\downarrow},\mathcal{K}(y_1)],\mathcal{K}(y_2)]+[\mathcal{K}(y_1),[Q_{\downarrow},\mathcal{K}(y_2)]].
\end{align}
We use,
\begin{align}
    [Q_{\downarrow},\mathcal{K}(y)]=\int_{-\infty}^{\infty}dx^- x^-[Q_{\downarrow},T_{--}(x^-,0,y)]=-\frac{i}{2}\int_{-\infty}^{\infty}dx^- x^-\partial_-\mathcal{J}_{-\downarrow}(x^-,0,y)=\frac{i}{2}q(y).
\end{align}
Therefore we obtain using the previous equation,
\begin{align}
   \mathfrak{f}(y_1,y_2)=[\mathcal{K}(y_1),q(y_2)]=[\mathcal{K}(y_2),q(y_1)]=\mathfrak{f}(y_2,y_1).
\end{align}
Next, we consider,
\begin{align}
    &[S_{\downarrow},[\mathcal{K}(y_1),\mathcal{K}(y_2)]]=0=[[S_{\downarrow},\mathcal{K}(y_1)],\mathcal{K}(y_2)]+[\mathcal{K}(y_1),[S_{\downarrow},\mathcal{K}(y_2)]]\notag\\
    &\implies y_1   \mathfrak{f}(y_1,y_2)=y_2   \mathfrak{f}(y_2,y_1)\implies (y_1-y_2)\mathfrak{f}(y_1,y_2)=0\notag\\
    &\implies \mathfrak{f}(y_1,y_2)=\delta(y_1-y_2)A_0(y_2).
\end{align}
Thus, we have ruled out the entire infinite tower of terms involving derivatives of delta functions and find,
\begin{align}
    [\mathcal{K}(y_1),q(y_2)]=\delta(y_1-y_2)A_0(y_2)=-\frac{i}{2}\delta(y_1-y_2)q(y_2),
\end{align}
where we fixed $A_0$ using the fact that $q(y)$ has boost eigen-value $\frac{1}{2}$. We can also easily determine $\{q(y_1),q(y_2)\}$ using this method. Construct,
\begin{align}
    &\{Q_{\downarrow},[\mathcal{K}(y_1),q(y_2)]\}=\{[Q_{\downarrow},\mathcal{K}(y_1)],q(y_2)]+[\mathcal{K}(y_1),\{Q_{\downarrow},q(y_2)\}]\notag\\
    &\implies -\frac{i}{2}\delta(y_1-y_2)[Q_{\downarrow},q(y_2)]=\frac{i}{2}\{q(y_1),q(y_2)\}+2[\mathcal{K}(y_1),\mathcal{E}(y_2)],\notag\\
    &\implies -i\delta(y_1-y_2)\mathcal{E}(y_2)=\frac{i}{2}\{q(y_1),q(y_2)\}-2i\delta(y_1-y_2)\mathcal{E}(y_2),\notag\\
    &\implies \{q(y_1),q(y_2)\}=2\delta(y_1-y_2)\mathcal{E}(y_2),
\end{align}
which is precisely the light-ray supersymmetry equation! Finally, let us determine $[\mathcal{N}_y(y_1),q(y_2)]$ this way. Consider the Jacobi identity,
\begin{align}
    &[\mathcal{N}_y(y_1),[Q_{\downarrow},\mathcal{K}(y_2)]]=[[\mathcal{N}_y(y_1),Q_{\downarrow}],\mathcal{K}(y_2)]+[Q_{\downarrow},[\mathcal{N}_y(y_1),\mathcal{K}(y_2)]],\notag\\
    &\implies \frac{i}{2}[\mathcal{N}_y(y_1),q(y_2)]=0+[Q_{\downarrow},-i\delta(y_1-y_2)\partial_{y_2}\mathcal{K}(y_2)+i\partial_{y_1}\delta(y_1-y_2)\mathcal{K}(y_2)]\notag\\
    &\implies [\mathcal{N}_y(y_1),q(y_2)]=-i\delta(y_1-y_2)\partial_{y_2}q(y_2)+i\partial_{y_1}\delta(y_1-y_2)q(y_2),
\end{align}
which is exactly the result we derived in the main-text.

\section{From $BMS_3$ to supersymmetric Virasoro}\label{app:bmsToVirasoro}
The $\mathfrak{bms}_3$ algebra can be realized as the symmetry algebra of asymptotically flat three dimensional spacetime. Similarly, the symmetry algebra preserving the structure of an asymptotically AdS$_3$ spacetime are two copies of the Virasoro algebra $\text{Vir}\otimes \text{Vir}$, as first shown by Brown and Henneaux. Similarly, when we have a supersymmetric bulk theory (with supersymmetry preserving boundary conditions) we expect a supersymmetric $\mathfrak{bms}_3$ algebra and a supersymmetric Virasoro algebra. In this appendix, we show that an appropriate cosmological constant deformation of the $\mathcal{N}=1$ $\mathfrak{bms}_3$ algebra results in the $\text{Super-Vir}\otimes \text{Vir}$ (which corresponds to $\mathcal{N}=(1,0)$ two dimensional super-conformal symmetry). This analysis can be considered as the reverse of the one performed in \cite{Banerjee:2016nio}. Consider the $\mathcal{N}=1$ $\mathfrak{bms}_3$ algebra \eqref{3dNeq1BMS} which we repeat here for convenience.
\begin{align}\label{3dNeq1BMS1}
   &[M_n,M_m]=0,\notag\\
    &[L_m,M_n]=(m-n)M_{m+n},\notag\\
    &[L_m,L_n]=(m-n)L_{m+n},\notag\\
     &\{F_r,F_s\}=2 M_{r+s},\notag\\
    &[F_r,M_n]=0,\notag\\
    &[F_r,L_m]=\bigg(r-\frac{m}{2}\bigg)F_{r+m}.
\end{align}
We consider turning on a negative cosmological constant $\Lambda=-\frac{1}{l^2}$ deformation which essentially makes the super-translations non-commutative. Below $\lambda=\frac{1}{l}$.
\begin{align}\label{3dNeq1BMS1deformedLambda}
   &[M_n,M_m]=\lambda^2(n-m)L_{n+m},\notag\\
    &[L_m,M_n]=(m-n)M_{m+n},\notag\\
    &[L_m,L_n]=(m-n)L_{m+n},\notag\\
     &\{F_r,F_s\}=2( M_{r+s}+\lambda L_{r+s}),\notag\\
    &[F_r,M_n]=\lambda\big(r-\frac{n}{2}\big)F_{r+n},\notag\\
    &[F_r,L_m]=\bigg(r-\frac{m}{2}\bigg)F_{r+m}.
\end{align}
This is a smooth deformation away from the $\Lambda=0$ flat spacetime. One can also check that the Jacobi identities are satisfied. Let us repackage the generators as follows:
\begin{align}
    \mathcal{L}_n=\frac{1}{2}(L_n+\frac{M_n}{\lambda}),\Bar{\mathcal{L}}_n=\frac{1}{2}(L_n-\frac{M_n}{\lambda}), Q_{r}=\frac{F_r}{2\sqrt{\lambda}}.
\end{align}
We find,
\begin{align}
    &[\mathcal{L}_n,\mathcal{L}_m]=\frac{1}{4}\bigg([L_n,L_m]+\frac{1}{\lambda}([L_n,M_m]+[M_n,L_m])+\frac{1}{\lambda^2}[M_n,M_m]\bigg)\notag\\
    &=\frac{(n-m)}{4}\bigg(L_{n+m}+\frac{2}{\lambda}M_{n+m}+\frac{1}{\lambda^2}\lambda^2L_{n+m}\bigg)=(n-m)\mathcal{L}_{n+m},\notag\\
    &[\mathcal{\Bar{L}}_n,\mathcal{\Bar{L}}_m]=\frac{1}{4}\bigg([L_n,L_m]-\frac{1}{\lambda}([L_n,M_m]-[M_n,L_m])+\frac{1}{\lambda^2}[M_n,M_m]\bigg)\notag\\
    &=\frac{(n-m)}{4}\bigg(L_{n+m}-\frac{2}{\lambda}M_{n+m}+\frac{1}{\lambda^2}\lambda^2L_{n+m}\bigg)=(n-m)\mathcal{\Bar{L}}_{n+m},\notag\\
    &[\mathcal{L}_n,\mathcal{\Bar{L}}_m]=\frac{1}{4}\bigg([L_n,L_m]+\frac{1}{\lambda}(-[L_n,M_m]+[M_n,L_m])-\frac{1}{\lambda^2}[M_n,M_m]\bigg)=0,
\end{align}
As for the re-defined super-supersymmetry generators,
\begin{align}
    &\{Q_r,Q_s\}=\frac{1}{4\lambda}\{F_r,F_s\}=\frac{1}{2}(L_{r+s}+\frac{M_{r+s}}{\lambda})=\mathcal{L}_{r+s}.
\end{align}
 Thus we find,
\begin{align}
    &[\mathcal{L}_m,\mathcal{L}_n]=(m-n)\mathcal{L}_{m+n},\{Q_r,Q_s\}=\mathcal{L}_{r+s},\notag\\
    &[\mathcal{\Bar{L}}_m,\mathcal{\Bar{L}}_n]=(m-n)\mathcal{\Bar{L}}_{m+n},\notag\\
    &[\mathcal{L}_n,\mathcal{\Bar{L}}_m]=0.
\end{align}
which is indeed the symmetry algebra of a $\mathcal{N}=(1,0)$ two dimensional conformal field theory with a left-handed super-Virasoro algebra and a right handed Virasoro algebra (with zero central charge).

\bibliographystyle{JHEP}
\bibliography{biblio}
\end{document}